\documentclass[ALICE,manyauthors]{cernphprep}
\usepackage[comma,square,numbers,sort&compress]{natbib}
\usepackage{hyperref}
\usepackage{lineno}
\usepackage{xcolor}
\usepackage{xspace}
\usepackage[mathscr]{eucal}
\usepackage[Symbolsmallscale]{upgreek}
\usepackage{multirow}
\usepackage{amsmath}
\usepackage{tabularx}
\usepackage{adjustbox}
\usepackage{extarrows}
\usepackage[T1]{fontenc}
\usepackage{orcidlink}

\begin{document}
%%%%%%%%%%%%%%%%%%%%%%%%%%%%%%%%%%%%%%%%%%%%%%%%%%
% These are some new commands that may be useful 
% for paper writing in general. If other newcommands
% are needed for your specific paper, please feel 
% free to add here. 
%
% The currently available commands are organized in: 
% 1) Systems
% 2) Quantities
% 3) Energies and units
% 4) Detectors
% 5) particle species 
%%%%%%%%%%%%%%%%%%%%%%%%%%%%%%%%%%%%%%%%%%%%%%%%%%

% 1) SYSTEMS 
\newcommand{\pp}           {pp\xspace}
\newcommand{\ppbar}        {\mbox{$\mathrm {p\overline{p}}$}\xspace}
\newcommand{\XeXe}         {\mbox{Xe--Xe}\xspace}
\newcommand{\PbPb}         {\mbox{Pb--Pb}\xspace}
\newcommand{\pA}           {\mbox{pA}\xspace}
\newcommand{\pPb}          {\mbox{p--Pb}\xspace}
\newcommand{\AuAu}         {\mbox{Au--Au}\xspace}
\newcommand{\dAu}          {\mbox{d--Au}\xspace}

% 2) QUANTITIES 
\newcommand{\s}            {\ensuremath{\sqrt{s}}\xspace}
\newcommand{\snn}          {\ensuremath{\sqrt{s_{\mathrm{NN}}}}\xspace}
\newcommand{\pt}           {\ensuremath{p_{\rm T}}\xspace}
\newcommand{\meanpt}       {$\langle p_{\mathrm{T}}\rangle$\xspace}
\newcommand{\ycms}         {\ensuremath{y_{\rm CMS}}\xspace}
\newcommand{\ylab}         {\ensuremath{y_{\rm lab}}\xspace}
\newcommand{\etarange}[1]  {\mbox{$\left | \eta \right |~<~#1$}}
\newcommand{\yrange}[1]    {\mbox{$\left | y \right |~<~#1$}}
\newcommand{\dndy}         {\ensuremath{\mathrm{d}N_\mathrm{ch}/\mathrm{d}y}\xspace}
\newcommand{\dndeta}       {\ensuremath{\mathrm{d}N_\mathrm{ch}/\mathrm{d}\eta}\xspace}
\newcommand{\avdndeta}     {\ensuremath{\langle\dndeta\rangle}\xspace}
\newcommand{\dNdy}         {\ensuremath{\mathrm{d}N_\mathrm{ch}/\mathrm{d}y}\xspace}
\newcommand{\Npart}        {\ensuremath{N_\mathrm{part}}\xspace}
\newcommand{\Ncoll}        {\ensuremath{N_\mathrm{coll}}\xspace}
\newcommand{\dEdx}         {\ensuremath{\textrm{d}E/\textrm{d}x}\xspace}
\newcommand{\RpPb}         {\ensuremath{R_{\rm pPb}}\xspace}

% 3) ENERGIES, UNITS
\newcommand{\nineH}        {$\sqrt{s}~=~0.9$~Te\kern-.1emV\xspace}
\newcommand{\seven}        {$\sqrt{s}~=~7$~Te\kern-.1emV\xspace}
\newcommand{\twoH}         {$\sqrt{s}~=~0.2$~Te\kern-.1emV\xspace}
\newcommand{\twosevensix}  {$\sqrt{s}~=~2.76$~Te\kern-.1emV\xspace}
\newcommand{\five}         {$\sqrt{s}~=~5.02$~Te\kern-.1emV\xspace}
\newcommand{\twosevensixnn}{$\sqrt{s_{\mathrm{NN}}}~=~2.76$~Te\kern-.1emV\xspace}
\newcommand{\fivenn}       {$\sqrt{s_{\mathrm{NN}}}~=~5.02$~Te\kern-.1emV\xspace}
\newcommand{\onethree}        {$\sqrt{s}~=~13$~Te\kern-.1emV\xspace}
\newcommand{\LT}           {L{\'e}vy-Tsallis\xspace}
\newcommand{\GeVc}         {Ge\kern-.1emV/$c$\xspace}
\newcommand{\MeVc}         {Me\kern-.1emV/$c$\xspace}
\newcommand{\TeV}          {Te\kern-.1emV\xspace}
\newcommand{\GeV}          {Ge\kern-.1emV\xspace}
\newcommand{\MeV}          {Me\kern-.1emV\xspace}
\newcommand{\GeVmass}      {Ge\kern-.1emV/$c^2$\xspace}
\newcommand{\MeVmass}      {Me\kern-.1emV/$c^2$\xspace}
\newcommand{\lumi}         {\ensuremath{\mathcal{L}}\xspace}

% 4) DETECTORS 
\newcommand{\ITS}          {\rm{ITS}\xspace}
\newcommand{\TOF}          {\rm{TOF}\xspace}
\newcommand{\ZDC}          {\rm{ZDC}\xspace}
\newcommand{\ZDCs}         {\rm{ZDCs}\xspace}
\newcommand{\ZNA}          {\rm{ZNA}\xspace}
\newcommand{\ZNC}          {\rm{ZNC}\xspace}
\newcommand{\SPD}          {\rm{SPD}\xspace}
\newcommand{\SDD}          {\rm{SDD}\xspace}
\newcommand{\SSD}          {\rm{SSD}\xspace}
\newcommand{\TPC}          {\rm{TPC}\xspace}
\newcommand{\TRD}          {\rm{TRD}\xspace}
\newcommand{\VZERO}        {\rm{V0}\xspace}
\newcommand{\VZEROA}       {\rm{V0A}\xspace}
\newcommand{\VZEROC}       {\rm{V0C}\xspace}
\newcommand{\Vdecay} 	   {\ensuremath{V^{0}}\xspace}

% 4) PARTICLE SPECIES 
\newcommand{\ee}           {\ensuremath{e^{+}e^{-}}} 
\newcommand{\pip}          {\ensuremath{\pi^{+}}\xspace}
\newcommand{\pim}          {\ensuremath{\pi^{-}}\xspace}
\newcommand{\pio}          {\ensuremath{\pi}\xspace} 
\newcommand{\kap}          {\ensuremath{\rm{K}^{+}}\xspace}
\newcommand{\kam}          {\ensuremath{\rm{K}^{-}}\xspace}
\newcommand{\kp}           {\ensuremath{\rm{K}^{+}}}
\newcommand{\km}           {\ensuremath{\rm{K}^{-}}}
\newcommand{\kpm}          {\ensuremath{\rm{K}^{\pm}}\xspace}
\newcommand{\kbar}         {\ensuremath{\rm\overline{K}}\xspace}
\newcommand{\pbar}         {\ensuremath{\rm\overline{p}}\xspace}
\newcommand{\kzeroS}       {\ensuremath{{\rm K}^{0}_{\rm{S}}}\xspace}
\newcommand{\kzero}        {\ensuremath{{\rm K}^{0}}\xspace}
\newcommand{\kzerobar}     {\ensuremath{\rm \overline{K}^0}}
\newcommand{\lmb}          {\ensuremath{\Lambda}\xspace}
\newcommand{\almb}         {\ensuremath{\overline{\Lambda}}\xspace}
\newcommand{\Om}           {\ensuremath{\Omega^-}\xspace}
\newcommand{\Mo}           {\ensuremath{\overline{\Omega}^+}\xspace}
\newcommand{\X}            {\ensuremath{\Xi^-}\xspace}
\newcommand{\Ix}           {\ensuremath{\overline{\Xi}^+}\xspace}
\newcommand{\Xis}          {\ensuremath{\Xi^{\pm}}\xspace}
\newcommand{\Oms}          {\ensuremath{\Omega^{\pm}}\xspace}
\newcommand{\degree}       {\ensuremath{^{\rm o}}\xspace}

%%Interaction pairs
\newcommand{\kbarN}         {\ensuremath{\rm\overline{K}N}\xspace}
\newcommand{\LPi}         {\ensuremath{\uppi\Lambda}\xspace}
\newcommand{\SPi}         {\ensuremath{\uppi\Sigma}\xspace}
\newcommand{\SipPim}           {\ensuremath{\rm \uppi^-\Sigma^+}\xspace}
\newcommand{\SimPip}           {\ensuremath{\rm \uppi^+\Sigma^-}\xspace}
\newcommand{\SizPiz}           {\ensuremath{\rm \uppi^0\Sigma^0}\xspace}
\newcommand{\LiPiz}           {\ensuremath{\rm \uppi^0\Lambda}\xspace}
\newcommand{\pXim}          {\ensuremath{\rm p\Xi^-}\xspace}
\newcommand{\pOm}           {\ensuremath{\rm p\Omega^-}\xspace}

%%%
%%%  Alias
%%%
\newcommand{\dedx}   {d$E$/d$x$}
\newcommand{\mom}    {\mbox{\rm MeV$\kern-0.15em /\kern-0.12em c$}}
\newcommand{\gmom}   {\mbox{\rm GeV$\kern-0.15em /\kern-0.12em c$}}
\newcommand{\mass}   {\mbox{\rm GeV$\kern-0.15em /\kern-0.12em c^2$}}
\newcommand{\Mmass}  {\mbox{\rm MeV$\kern-0.15em /\kern-0.12em c^2$}}
\newcommand{\kstar}  {\ensuremath{k^{*}}\xspace}
\newcommand{\ap}     {$\overline{\mathrm{p}}$}
\newcommand{\spe}    {$S_\mathrm{T}$\xspace}
\newcommand{\opl}    {$\oplus$}
\newcommand{\antik}   {$\mathrm{\overline{K}}\,$} 
\newcommand{\antikz}   {$\mathrm{\overline{K}^0}$}
\newcommand{\rcore}  {\ensuremath{r_{\rm core}}\xspace}
\newcommand{\temp}  {\ensuremath{T_{\rm ch}}\xspace}

\newcommand{\Ledn}         {Lednick\'y--Lyuboshits\xspace}
\newcommand{\chiEFT}       {\ensuremath{\chi}\rm{EFT}\xspace}
%CORRELATION FUNCTIONS
\newcommand{\ks}     {\ensuremath{k^{*}}\xspace}
\newcommand{\rs}     {\ensuremath{r^{*}}\xspace}
\newcommand{\mt}     {\ensuremath{m_{\mathrm{T}}}\xspace}
\newcommand{\Cth}           {C_\mathrm{th}\xspace}
\newcommand{\Cexp}           {C_\mathrm{exp}\xspace}
\newcommand{\CF}           {\ensuremath{C(\ks)}\xspace}
\newcommand{\Sr}            {\ensuremath{S(\rs)}\xspace}

%%%%%%%%%%%%%%%  Title page %%%%%%%%%%%%%%%%%%%%%%%%
\begin{titlepage}
% the dates below correspond to CERN approval
% please don't touch: EB chairs will take care
\PHyear{2022}       % required, will be obtained from CERN
\PHnumber{107}      % required, will be obtained from CERN
\PHdate{25 May}  % required, will be obtained from CERN
%%%%%%%%%%%%%%%%%%%%%%%%%%%%%%%%%%%%%%%%%%%%%%%%%%%%

%%% Put your own title + short title here:
\title{Constraining the \kbarN coupled channel dynamics using femtoscopic correlations at the LHC}
\ShortTitle{Constraining the \kbarN coupled channels}   % appears on left page headers

%%% Do not change the next lines
\Collaboration{ALICE Collaboration\thanks{See Appendix~\ref{app:collab} for the list of collaboration members}}
\ShortAuthor{ALICE Collaboration} % appears on right page headers, do not change

\begin{abstract}
The interaction of \km with protons is characterised by the presence of several coupled channels, systems like \kzerobar n and \SPi with a similar mass and the same quantum numbers as the \km p state. The strengths of these couplings to the \km p system are of crucial importance for the understanding of the nature of the $\Lambda(1405)$ resonance and of the attractive \km p strong interaction. In this article, we present measurements of the \km p correlation functions in relative momentum space obtained in \pp collisions at \onethree, in  \pPb collisions at \fivenn, and (semi)peripheral \PbPb collisions at \fivenn. The emitting source size, composed of a core radius anchored to the \kp p correlation and of a resonance halo specific to each particle pair, varies between 1 and 2~fm in these collision systems. The strength and the effects of the \kzerobar n and \SPi inelastic channels on the measured \km p correlation function are investigated in the different colliding systems by comparing the data with state-of-the-art models of chiral potentials. A novel approach to determine the conversion weights $\omega$, necessary to quantify the amount of produced inelastic channels in the correlation function, is presented. In this method, particle yields are estimated from thermal model predictions, and their kinematic distribution from blast-wave fits to measured data.
The comparison of chiral potentials to the measured \km p interaction indicates that, while the \mbox{\SPi--\km p} dynamics is well reproduced by the model, the coupling to the \kzerobar n channel in the model is currently underestimated.

\end{abstract}
\end{titlepage}

\setcounter{page}{2} %please do not remove this line

%%%%%%%%%%%%%%%%%%%%%%%%%%%%%%%%
% begin main text
%%%%%%%%%%%%%%%%%%%%%%%%%%%%%%%%
\section{Introduction}\label{sec:Intro}
The interaction between antikaons (\kbar) and nucleons (N) is one of the main building blocks of \mbox{low-energy} effective field theories aiming at describing the non-perturbative regime of the strong interaction with strangeness degrees of freedom.
The \kbarN interaction is characterised by the presence of several coupled channels, such as \LiPiz and \SPi~\cite{Dalitz:1959dn,Dalitz:1960du}, with the same quantum numbers as \kbarN and an invariant mass below that of the \kbarN system~\cite{Hyodo:2011ur,Meissner:2020khl,Mai:2020ltx,Hyodo:2020czb} (i.e. sub-threshold). This coupled channel dynamics is responsible for the inelastic component of the \kbarN interaction, accounting for transitions like $\kbarN \xleftrightarrow{}  \LiPiz,\SPi$~\cite{Haidenbauer:2018jvl,Kamiya:2019uiw}. 

While the coupling of \kbarN to \LiPiz is negligible~\cite{LPiCoupling1,LPiCoupling2}, the one to \SPi is dominant in the sub-threshold region.
At low energy, the dynamics between the \kbarN and the \SPi channel leads to the formation of the $\Lambda(1405)$ resonance, approximately 27 MeV below the threshold. The internal structure of the $\Lambda(1405)$ has been widely investigated and currently it is the only accepted molecular state arising from the interplay of the \mbox{\kbarN--\SPi} poles~\cite{Hall:2014uca,Kamiya:2015aea,Kamiya:2016oao}.
Measurements of the $\Lambda(1405)$ spectral shape in the strong decays to the \SPi final state can provide information on the \mbox{\kbarN--\SPi} coupling, but direct access is mainly hampered by the limited amount of data, by the presence of interference terms in the total amplitude due to baryonic resonances in the final state~\cite{Nacher:1998mi,Lambda1405_1,Lambda1405_2,Lambda1405_3,Lambda1405_4,Lambda1405_CLAS}, and ultimately by differences in the theoretical description of the \kbarN interaction below threshold. In fact, the behaviours of the \kbarN interaction and of the coupling to the \SPi channel in the sub-threshold region are anchored to the knowledge of such interaction above and at the threshold. The measurement of kaonic hydrogen~\cite{Bazzi:2011zj} provides the most precise constraint on the \kbarN interaction at the threshold, but the coupled channel dynamics is included in the extracted \kbarN scattering parameters only indirectly~\cite{Hyodo:2011ur}.

Scattering experiments available above the \kbarN threshold, in which the initial state (typically \km p) is fixed~\cite{Humphrey:1962zz, Watson:1963zz, Mast:1975pv, Nowak:1978au, Ciborowski:1982et}, represent the largest source of constraints for the available theoretical approaches~\mbox{\cite{KAISER1995325,OSET199899,OLLER2001263,LUTZ2002193,Hyodo:2007jq,KAMIYA201641,Revai:2017isg,Mai:2014xna,Borasoy:2006sr,Cieply:2015pwa}}. Access to the \kbarN interaction and to the different inelastic couplings is achieved by measurements of elastic and inelastic cross sections of the different final states. 
In the specific case of the \km p system, another inelastic channel, \kzerobar n, appears approximately 5~MeV above the \km p threshold, due to the mass difference between neutral kaons and neutrons and their charged isospin partners (\km, p). A cusp structure in the total \km p cross section is predicted to occur at the opening of this coupled channel~\cite{Ikeda:2012au}, but no evidence in scattering data has been observed yet due to the insufficient statistical precision~\mbox{\cite{Mast:1975pv,Ciborowski:1982et,Sakitt:1965kh}}. As a consequence, the theoretical description of the dynamics between the \km p and \kzerobar n channels is currently not fully constrained by experimental data.

Recently, this scenario changed due to the measurements of correlations of (\km p \opl\ \kp \ap) (for brevity \km p) pairs in \pp collisions at centre-of-mass energies of \five, 7~TeV and 13~TeV by the ALICE Collaboration, which provided the first experimental evidence for the opening of the \kzerobar n channel and the most precise data on the \km p interaction down to zero relative momenta~\cite{FemtoKp_pp}. 
The femtoscopy technique~\cite{Lednicky:2005af,Lisa:2005dd} applied in small colliding systems, in which particles are emitted by a source with typical size of about 1~fm, can easily access the short-range part of the strong interaction in which the coupled channel dynamics plays an important role. The measured correlation functions show great sensitivity to the underlying strong potential~\cite{ALICE:Run1,ALICE:LL,ALICE:pSig0,ALICE:pXi,ALICE:pOmega} and to the different inelastic channels, affecting both the shape and magnitude of the femtoscopic signal~\cite{Fabbietti:2020bfg,Kamiya:2019uiw,Haidenbauer:2018jvl}. 
The recent results by the ALICE Collaboration on \km p pairs in central \PbPb collisions showed quantitatively that the contributions of (\kzerobar n \opl\ K$^0\overline{\mathrm{n}}$) (hereafter \kzerobar n) and \SPi channels are negligible for inter-particle distances of 5 fm and above~\cite{FemtoKp_PbPb}. The results obtained in \pp and \PbPb collisions clearly suggest the possibility to constrain the \km p interaction and the different inelastic coupling by measuring \km p pairs in different colliding systems.

In this article, the femtoscopic measurements of (\kp p \opl\ \km \ap) (hereafter \kp p) and \km p 
pairs obtained in three different centrality intervals in \pPb collisions at collision energy per nucleon pair \fivenn and in three different centrality intervals in peripheral \PbPb collisions at \fivenn are presented. Different centrality intervals were chosen to probe different radii of the particle emitting source. Results from the reanalysis of the femtoscopic data obtained in minimum bias \pp collisions at \onethree shown in Ref.~\cite{FemtoKp_pp} are also included. In this study, the data from Ref.~\cite{FemtoKp_pp} are corrected for the finite experimental momentum resolution and analysed again, employing the same  procedure used for the other colliding systems.
The same-charge pairs, \kp p, for which the interaction is well known
and no inelastic channels are present~\cite{Hadjimichef:2002xe,AokiJidoKplusp}, are used as a benchmark to extract information on the emitting source size needed to evaluate the correlation function. The measured correlations of \km p 
are compared with state-of-the-art chiral potentials~\cite{Kamiya:2019uiw,Miyahara:2018onh} derived in a coupled channel approach. A detailed investigation on the strong coupling to \SPi and \kzerobar n channels is performed in the three different colliding systems.

This paper is organised as follows: a short description of the ALICE detector can be found in Section~\ref{sec:alice} while in Section~\ref{sec:Dataanalysis} the event and track selection criteria are discussed together with the particle identification technique and the systematic uncertainties evaluation on the correlation function. In Section~\ref{sec:CFanalysis}, the analysis and the modeling of the correlation function are introduced and the determination of the emitting source size is described; the obtained results are presented and discussed in Section~\ref{sec:Results}. Finally, in Section~\ref{sec:Summary}, conclusions and future outlooks are provided.

\section{The ALICE detector}
\label{sec:alice}
A detailed description of the ALICE experimental setup can be found 
in Refs.~\cite{ALICE:2008ngc,Abelev:2014ffa} and references therein.
The main sub-detectors used in this analysis are: the V0 detectors~\cite{Abbas:2013VZERO}, the Inner Tracking System (ITS)~\cite{Aamodt:2010aa}, 
the Time Projection Chamber (TPC)~\cite{Alme:2010TPC} and the Time-Of-Flight (TOF) detector
\cite{Akindinov:2013tea}. The ITS, TPC and TOF are located inside a solenoidal magnet that provides a uniform field of 0.5~T directed along the beam direction. All the detectors in the central barrel region used for this analysis (ITS, TPC and TOF) cover the full azimuth and have a pseudorapidity coverage of $|\eta|<0.9$. 

The V0 detector consists of two arrays of scintillation counters placed on either side of the interaction point: one covering the pseudorapidity interval $2.8 < \eta < 5.1$~\mbox{(V0A)} and the other one covering \mbox{$-3.7 < \eta < -1.7$}~\mbox{(V0C)}. 
The scintillator arrays have an intrinsic time resolution better than 0.5~ns. Their timing information is used in coincidence for offline rejection of events produced by the interaction of the beams with residual gas in the vacuum pipe. 
The V0 scintillators are used to determine the collision centrality from the measured charged-particle multiplicity~\cite{Aamodt:2011oai, Abelev:2013qoq, CentralityDeterminationPublicNote}.  

The ITS, designed to provide high-resolution track points close to the beam line, is composed of three subsystems of silicon detectors placed around the interaction region with a cylindrical symmetry. The Silicon Pixel Detector (SPD) is the closest subsystem to the beam pipe and is made of two layers of pixel detectors. The third and the fourth layers are formed by silicon drift detectors, while the outermost two layers are equipped with double-sided silicon strip detectors. 

The TPC, which is the main tracking detector, consists of a hollow cylinder whose axis coincides with the nominal beam axis and which surrounds the
ITS. 
The charged-particle tracks are then built by combining the hits in the ITS and up to 159 reconstructed space points in the TPC. The momentum component transverse to the beam pipe (\pt) of charged particles is reconstructed from the curvature of the tracks within the TPC, which is permeated by a magnetic field.
The TPC is also used for particle identification (PID) via the measurement of the specific energy loss (\dedx) in its gas volume.

The TOF detector is based on the multi-gap resistive plate chambers technology and is located, with a cylindrical symmetry, at an average radial distance of 380~cm from the beam axis. The particle velocity $\beta$ of each particle can be determined by using the time-of-flight measurement, the track length, and the associated momentum, allowing for PID at intermediate momenta.  

\section{Event and track selection}\label{sec:Dataanalysis}

The data samples used for the measurements presented in this paper were recorded by ALICE in 2015, 2016, 2017, and 2018 during the LHC pp runs at \onethree, the \pPb runs at \fivenn and the \PbPb runs at \fivenn. A minimum bias trigger was used during all the data taking, which required coincident signals in both V0 scintillators to be synchronous with the beam crossing time defined by the LHC clock. Events with multiple primary vertices identified with the SPD are tagged as pile-up and excluded from the analysis to achieve the best PID performance.

The primary vertex position was determined from tracks reconstructed in the ITS and TPC as described in Ref.~\cite{Abelev:2014ffa} and only events with a reconstructed primary vertex position along the beam axis  ($V_z$) within 10~cm from the nominal interaction point along the beam direction are selected. This procedure ensures a uniform detector coverage within $|\eta|<0.8$. After the application of the event selection criteria, about 1$\times10^9$ minimum bias pp events, about 8$\times10^8$ minimum bias \pPb collisions in the 0--100\% multiplicity interval and about 6.5$\times10^7$ minimum bias \PbPb\ collisions in the 
60--90\% centrality interval, which corresponds to most peripheral events, have been analysed. 

Small collision systems are affected by the presence of mini-jets background,  which might affect the femtoscopic signal~\cite{FemtoKp_pp,ALICE:BBarpp}. To reduce the contribution from the mini-jet background, the events were classified according to their transverse sphericity (\spe), an observable which is known to be correlated with the number of hard parton--parton interactions in each event~\cite{Abelev:2012sk}. An event with hard parton--parton interactions will generally produce a jet-like event topology that yields low sphericity, while an event dominated by soft parton--parton interactions can yield higher sphericity. 
To reduce the strong mini-jet background at low momenta, only events 
with \spe, defined as in Ref.~\cite{Acharya:2019idg}, larger than 0.7 were considered in the pp analysis. Generally, the mini-jet background is most pronounced in pp while in \pPb and peripheral \PbPb the events are typically isotropic.  However, for consistency, the same selection on \spe was applied to the \pPb and semi-peripheral \PbPb data. 

Charged kaon and proton candidate tracks were selected from charged-particle tracks reconstructed in the TPC, with the additional constraint that the track originates from the primary vertex~\cite{Abelev:2014ffa}. Candidate tracks were selected in the  range $|\eta|<~$0.8. 
To assure a good \pt resolution and to remove wrongly reconstructed tracks from the sample, only tracks with at least 70 space points %out of a maximum of 159
were selected. To suppress contributions of secondary kaons and protons in the sample, the reconstructed tracks were required to have a distance of closest approach to the primary vertex (DCA) along both the beam ($z$) and transverse ($x,y$) directions smaller than 1~cm.  In order to reduce the hadron misidentification, kaon candidates with 0.15~$<$~\pt~$<$~1.4~\gmom\ and proton candidates with 0.4~$<$~\pt~$<$~3~\gmom\ were selected.

For particle identification, both the TPC and the TOF detectors were employed. Kaons (protons) with transverse momenta up to 0.4 (0.8)~\gmom\ were identified using only the TPC information by requiring that the average \dEdx\ (calculated using the Bethe--Bloch parameterisation) is within three standard deviations ($\sigma$) from the expected average at a given momentum for the particle mass hypothesis. 
For kaons with \pt~$>$~0.4~\gmom\ and protons with \pt~$>$~0.8~\gmom, a similar method was applied for the particle identification using the TOF, where, on top of TPC selection, a 3$\sigma$ selection on the expected time of flight for a given particle at a given momentum was applied. Tracks for which the PID was ambiguous were discarded. 
To remove the large fraction of e$^+$e$^-$ pairs that can affect the extraction of the correlation function of the opposite-charge pairs, a selection on the \pt\ of kaons and protons of each charge was applied: kaon candidates were excluded if \mbox{0.3~$<$\pt~$<$~0.4~\gmom}, while proton candidates 
were excluded in the interval between 0.6~$<$~\pt~$<$~0.8~\gmom. It should be noted that cutting in this \pt region does not influence the correlation signal as the statistical uncertainty of the data is sufficiently small in each investigated \ks interval. The quoted \pt intervals correspond to the regions in which the electron \dEdx band merges with the band of the kaons and protons, respectively. The purity of the selected particle samples was determined by Monte Carlo simulations based on PYTHIA~8~\cite{Sjostrand:2007gs} for the pp
analysis, on EPOS~\cite{EPOS} for \pPb, and on HIJING~\cite{Wang:1991h} for \PbPb. The purity is larger than 99\% in the considered \pt\ intervals for all the analysed data. 

\section{Analysis of the correlation function}\label{sec:CFanalysis}

\subsection{Experimental correlation function}\label{sec:ck_exp}
The main observable in this work is the two-particle correlation function. Experimentally, the correlation function $C(\ks)^{\rm measured}$ is constructed as~\cite{Lisa:2005dd}

\begin{align}\label{eq:CFexp}
 & C(\ks)^{\rm measured} = \mathcal{N} \dfrac{A(\ks)}{B(\ks)}, 
\end{align}
where \ks  is the magnitude of the momentum of each of the particles in the pair reference frame.
The numerator $A$(\kstar) is the measured distribution of pairs from the same event, while $B$(\kstar) is the reference distribution of pairs from mixed events. 
The pairs in the denominator are formed by mixing particles from one event with particles from a  pool of up to seven other events with a comparable number of charged particles at midrapidity~\cite{ALICE-PUBLIC-2019-006}, a distance between primary vertex coordinate $V_z$ along the beam axis \mbox{$\Delta V_z \leq$ 2~cm} and with similar \spe as for the numerator. 

The $\mathcal{N}$ 
parameter is chosen such that the mean value of the correlation function equals unity for 600~$<$~\kstar~$<$~1000~\mom\,where the correlation function is flat. Reconstruction biases, such as the merging of tracks that are very close to each other, were evaluated by means of Monte Carlo simulations and found to be negligible in all the collision systems and centralities considered in this analysis. The measured \kstar is not identical to the true relative momentum of the pair due to effects of momentum resolution~\cite{ALICE:pL}. Hence, to compare the experimental results with theoretical predictions, an unfolding of the data is required. This was done by applying a Bayesian unfolding method~\cite{Adye:2011gm} both to the to the $A$(\kstar) and $B$(\kstar) distributions. The finite experimental momentum resolution modifies the measured correlation functions at most by 6\% in the first \kstar interval and less than 1\% for \kstar around 300~\mom. The resulting $C(\ks)^{\rm corrected}$ correlation function can be described by 

\begin{equation}
    C(\kstar)^{\rm corrected} = (a + b \, k^*) \left( 1+ \lambda_{ \mathrm{genuine}} (C(\kstar)^ \mathrm{genuine}  - 1 )+ \sum_{ij} \lambda_{ij} \left( C_{ij} (\kstar) - 1 \right)\right).
    \label{eq:unfoldinge}
\end{equation}

The correlation function is dominated by the contribution of the genuine \kp p  or \km p interaction, $C(\kstar)^ \mathrm{genuine}$. It is weighted by the corresponding $\lambda_{ \mathrm{genuine}}$ parameter, which describes the purity and fractions of primary kaons and protons in the sample~\cite{ALICE:Run1}. 
Other contributions $i,j$ originating from incorrectly identified particles, from particles stemming from weak decays (such as protons from \mbox{$\Lambda \rightarrow \mathrm{p}\uppi^-$)}, and from particles coming from long-lived resonances (such as kaons from  $\phi$ (1020)$\rightarrow$\kp \km) give rise to the $C_{ij} (\kstar)$ correlation functions.
Since weak decays occur typically some centimetres away from the collision vertex, a negligible final-state interaction between their decay products and the primary particles under study can be assumed. A similar argument can be applied also to particles coming from long-lived resonances \mbox{($c\tau >  5\,\mathrm{fm}$)}. Hence, the resulting correlation functions  $C_{ij} (\kstar)$ originating from the different combinations of primary, secondary, and misidentified particles are considered independent of \ks. 
The correlations due to misidentifications are evaluated experimentally and their contribution depends on the fraction of primary pairs of the sample (around 70\% for kaon--proton pairs), on the fraction of $\phi$ (1020) that decay into kaon pairs and are identified as primary kaons (around 6\%), and on the purity of primary pairs in the analysed sample (around 98\% for kaon--proton pairs). These fractions are determined by fitting Monte Carlo templates to the measured DCA$_{xy}$ distributions of kaons and protons, similarly to what is described in Ref.~\cite{FemtoKp_pp}. 
All the contributions related to misidentified kaons and protons are encapsulated into the $\lambda_{ij}$ parameter of Eq.~(\ref{eq:unfoldinge}). 
Finally, an additional contribution related to energy-momentum and charge conservation is present in $C(\kstar)^{\rm corrected}$. This contribution is described by a linear baseline $(a+b \, \kstar)$ whose coefficients are fixed by fitting the measured correlation function in a \kstar region where the short-range effects are negligible. For \kp p the parameters are fixed in the 300~$<~\kstar~<$~400~\mom\ interval. In the \km p case, the baseline parameters are fixed using the data in the 180~$<~\kstar~<$~270~\mom\ region. 

Once all the above contributions have been taken into account, the genuine correlation $C(\kstar)^ \mathrm{genuine}$ (\CF from now on for simplicity) is obtained. 
The systematic uncertainties on \CF were evaluated for each \kstar\ interval by varying event and track selection criteria. 
The event sample was varied by changing the selection on the $V_z$ position from $\pm$10~cm to $\pm$7~cm and by varying the sphericity of the accepted events from \spe~$>$~0.7 to  \spe~$>$~0.6 and \spe~$>$~0.8.  
Systematic uncertainties related to the track selection criteria were studied by varying the selection on the DCA$_{xy}$ within the experimental resolution. Systematic uncertainties related to other variations related to track selections were found to be negligible. 
To study systematic effects related to particle identification, the number of standard 
deviations around the expected energy loss for kaons and protons in the TPC and, similarly, for 
the time of flight in the TOF detector was modified from 3$\sigma$ to 2$\sigma$. For each source, the systematic uncertainty was estimated as the root-mean-square (RMS) 
of all the deviations in each \kstar interval. The total systematic uncertainty was  calculated as the square root of the quadratic sum of 
each source's contribution and amounts to about 3\% up to \kstar~$<$~500~\mom. The uncertainties related to the correction procedure described in the previous paragraph are propagated into the measured data points by means of the bootstrap technique~\cite{Bootstrap}.

\begin{figure}[!htb]  
\begin{center}
\includegraphics[width=0.4\textwidth]{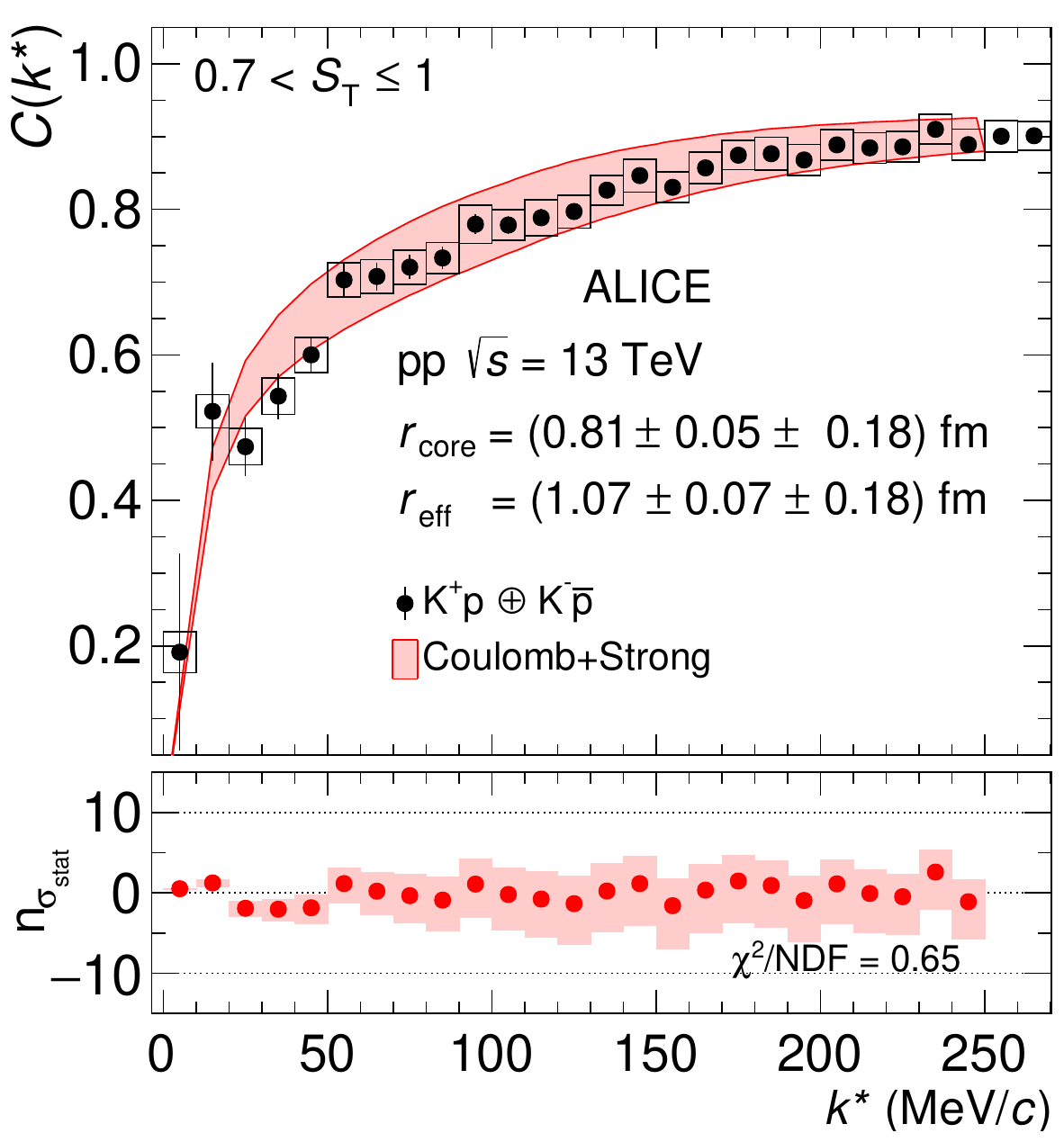} 
\end{center}
\caption{\kp p (\kp p \opl\ \km \ap) correlation function obtained in pp collisions at \onethree. The measured data points are taken from Ref.~\cite{FemtoKp_pp} and are corrected for finite experimental momentum resolution and for residual correlations as described in Section~\ref{sec:ck_exp}. Measured data are shown by the black markers, the vertical error bars and the boxes represent the statistical and systematic uncertainties, respectively. The red band in the upper panel represents the model calculation and its systematic uncertainty 
as described in the text. The $r_\mathrm{core}$ and $r_{\rm eff}$ values of the source are reported with their statistical and systematical uncertainties, respectively. Bottom panels represent the data-to-model comparison. }
\label{fig:KppCFfitpp}
\end{figure}

\begin{figure}[!htb]  
\begin{center}
\includegraphics[width=1\textwidth]{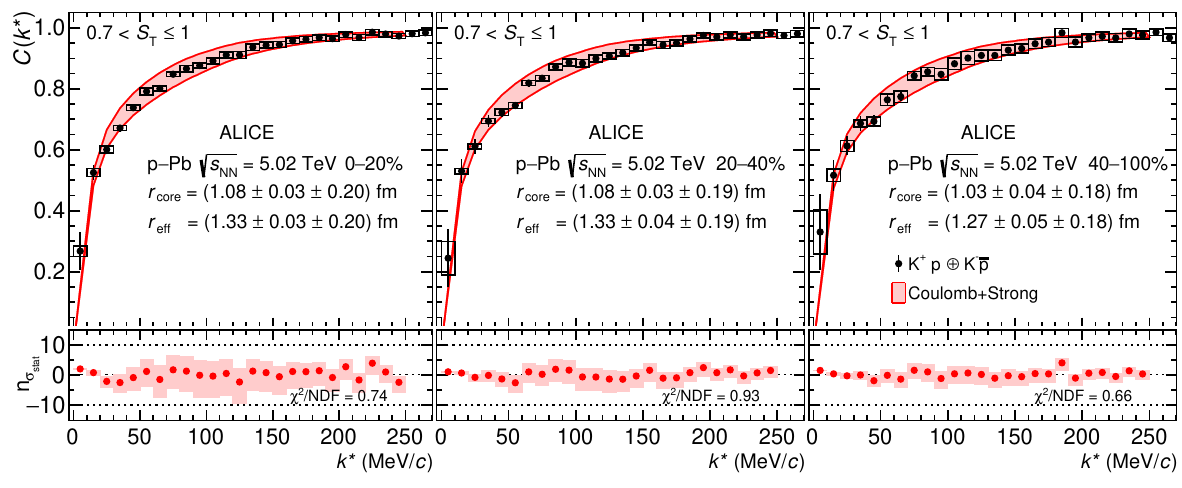}
\end{center}
\caption{\kp p (\kp p \opl\ \km \ap) correlation functions obtained in \pPb collisions at \fivenn
in the 0--20\% (left), 20--40\% (middle) and 40--100\% (right) centrality intervals. The measurement is shown by the black markers. The vertical error bars and the boxes represent the statistical and systematic uncertainties, respectively. The red band in the upper panels represents the model calculation and its systematic uncertainty  
as described in the text. The $r_\mathrm{core}$ and $r_{\rm eff}$ values of the source are reported with their statistical and systematical uncertainties, respectively. Bottom panels represent the data-to-model comparison.}
\label{fig:KppCFfitpPb}
\end{figure}

\begin{figure}[!htb]  
\begin{center}
\includegraphics[width=1\textwidth]{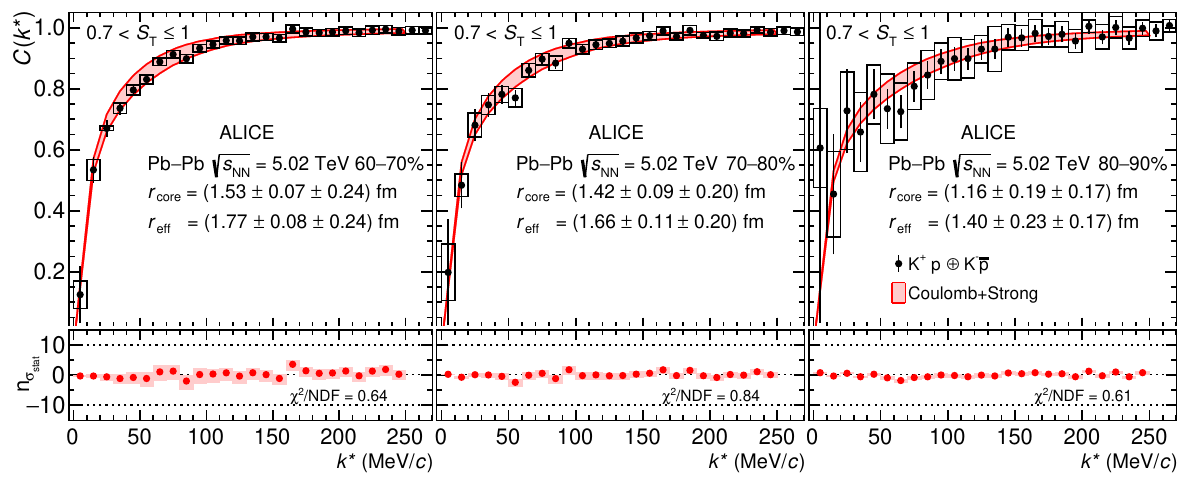}
\end{center}
\caption{\kp p (\kp p \opl\ \km \ap) correlation functions obtained in \PbPb collisions at \fivenn
in the 60--70\% (left), 70--80\% (middle) and 80--90\% (right) centrality intervals. The measurement is shown by the black markers. The vertical error bars and the boxes represent the statistical and systematic uncertainties, respectively. 
The red band in the upper panels represents the model calculation and its systematic uncertainty 
as described in the text. The $r_\mathrm{core}$ and $r_{\rm eff}$ values of the source are reported with their statistical and systematical uncertainties, respectively. Bottom panels represent the data-to-model comparison.}
\label{fig:KppCFfitPbPb}
\end{figure}

\begin{figure}[!htb]  
\begin{center}
\includegraphics[width=0.4\textwidth]{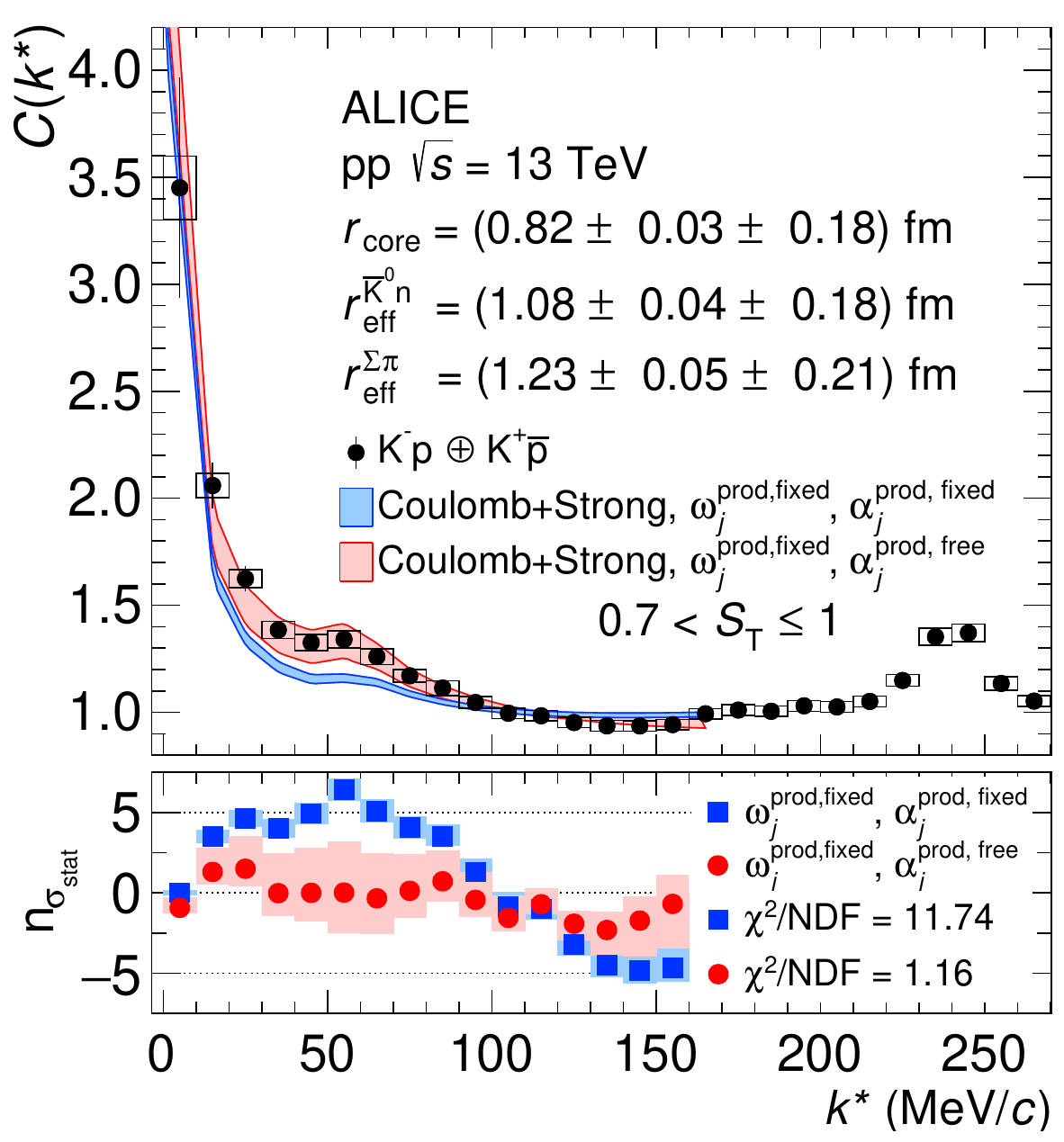} 
\end{center}
\caption{(\km p \opl\ \kp \ap) correlation functions obtained in pp collisions at \onethree. The measured data points are taken from~\cite{FemtoKp_pp} and have been corrected for finite experimental momentum resolution and for residual correlations as described in Section~\ref{sec:ck_exp}. Measured data are shown by the black markers, the vertical error bars and the boxes represent the statistical and systematic uncertainties, respectively. The red and blue bands in the upper panels represents the model calculations and its systematic uncertainty 
as described in the text. The $r_\mathrm{core}$ and $r_{\rm eff}$ values of the source are reported with their statistical and systematical uncertainties. Bottom panels represent the data-to-model comparison as described in the text.  }
\label{fig:KmpCFfitpp}
\end{figure}

\begin{figure}[!htb]  
\begin{center}
\includegraphics[width=1\textwidth]{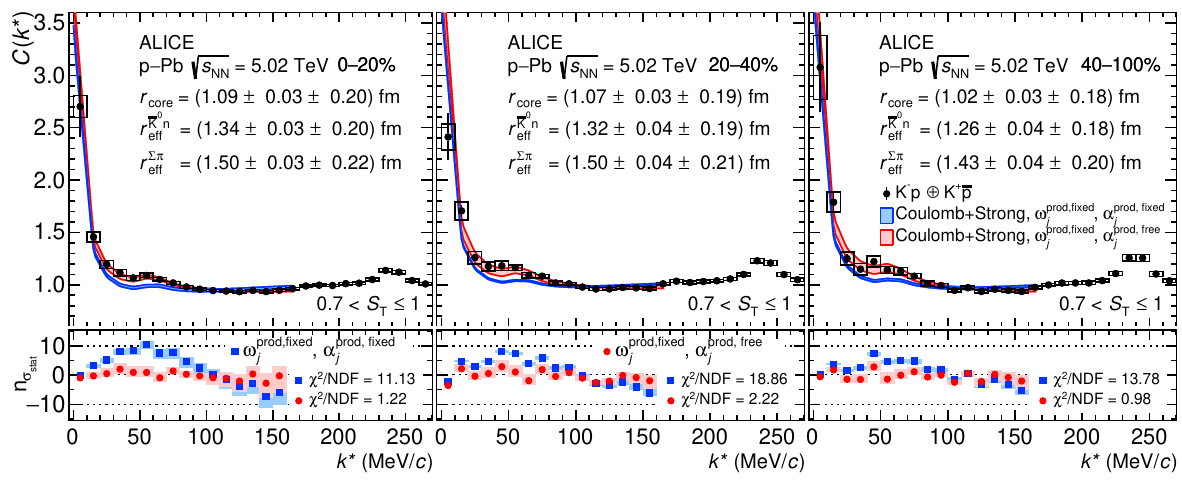}
\end{center}
\caption{(\km p \opl\ \kp \ap) correlation functions obtained in \pPb collisions at \fivenn
in the 0--20\% (left), 20--40\% (middle) and 40--100\% (right) centrality intervals. The measurement is shown by the black markers, the vertical error bars and the boxes represent the statistical and systematic uncertainties, respectively. The red and blue bands in the upper panels represent the model calculations and their systematic uncertainty 
as described in the text. The $r_\mathrm{core}$ and $r_{\rm eff}$ values of the source are reported with their statistical and systematical uncertainties, respectively. Bottom panels represent the data-to-model comparison as described in the text. 
}
\label{fig:KmpCFfitpPb}
\end{figure}

\begin{figure}[!htb]  
\begin{center}
\includegraphics[width=1\textwidth]{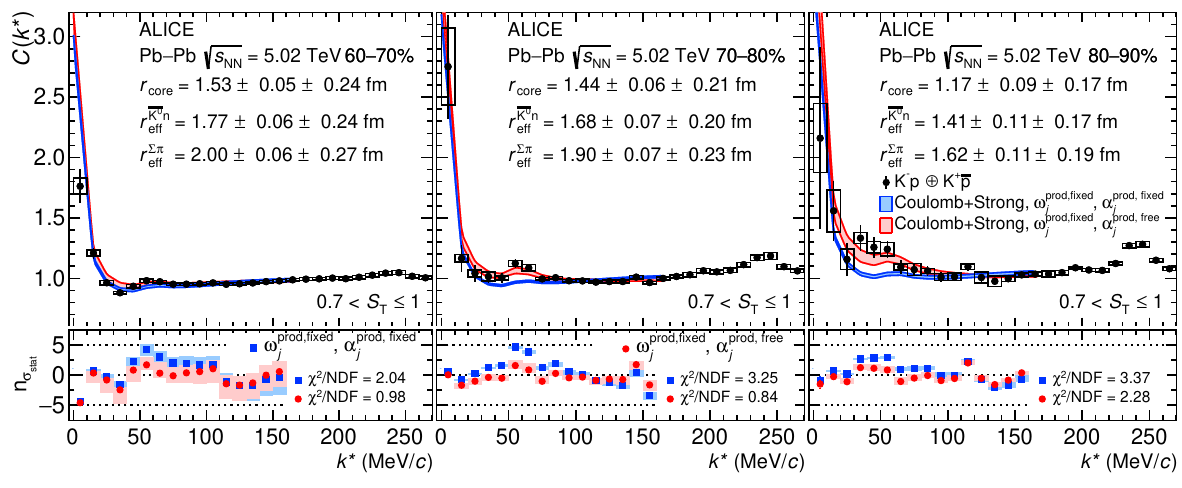}
\end{center}
\caption{(\km p \opl\ \kp \ap)  correlation functions obtained in \PbPb collisions at \fivenn
in the 60--70\% (left), 70--80\% (middle) and 80--90\% (right) centrality intervals. The measurement is shown by the black markers, the vertical error bars and the boxes represent the statistical and systematic uncertainties respectively. The red and blue bands in the upper panels represent the model calculations and their systematic uncertainty 
as described in the text. The $r_\mathrm{core}$ and $r_{\rm eff}$ values of the source are reported with their statistical and systematical uncertainties, respectively. Bottom panels represent the data-to-model comparison as described in the text. 
}
\label{fig:KmpCFfitPbPb}
\end{figure}

The \CF of \kp p are shown in Fig.~\ref{fig:KppCFfitpp} for pp collisions at \onethree and in Fig.~\ref{fig:KppCFfitpPb} and Fig.~\ref{fig:KppCFfitPbPb} for \pPb and \PbPb at \fivenn, respectively. In Fig.~\ref{fig:KppCFfitpPb} and Fig.~\ref{fig:KppCFfitPbPb}, each panel corresponds to a different centrality, as indicated in the legend of the figures. 
The data shown in Fig.~\ref{fig:KppCFfitpp}, already presented in~\cite{FemtoKp_pp}, are corrected as described above.  
Similarly, the \CF for \km p are shown in Fig.~\ref{fig:KmpCFfitpp} for \pp collisions, and in Fig.~\ref{fig:KmpCFfitpPb} and Fig.~\ref{fig:KmpCFfitPbPb} for \pPb at \fivenn and \PbPb at \fivenn, respectively. Also in this case the data obtained in pp collisions have already been shown in Ref.~\cite{FemtoKp_pp} and are reported here after the corrections described above. 

In Fig.~\ref{fig:KmpCFfitpp},  Fig.~\ref{fig:KmpCFfitpPb},  and Fig.~\ref{fig:KmpCFfitPbPb}, the structure that can be seen in the \km p 
correlation function at $k^*$ around 240~\mom\ is consistent with the $\Lambda$(1520) ($\overline{\Lambda}$(1520)) which decays into a \km p (\kp\ap) pair of relative momentum $k^* =243$~\mom~\cite{ParticleDataGroup:2020ssz}.
As already observed in pp collisions~\cite{FemtoKp_pp}, the correlation function for \km p 
exhibits a clear structure between 50 and 60 \mom\ visible for all the colliding systems, centralities and energies. The position of the structure at $\kstar=59$~\mom\ is consistent with the opening of the $\mathrm{\overline{K}^0n}$ 
channel corresponding to the kaon measured momentum in the laboratory frame  $p_{\mathrm{lab}}=89$~\mom~\mbox{\cite{Mast:1975pv,Ciborowski:1982et,Sakitt:1965kh}}.  

\subsection{Modeling the correlation function with inelastic channels}\label{sec:ck_theo}
For the modeling of the correlation function in the \km p system, the Koonin--Pratt formula~\cite{Lisa:2005dd} is modified to take into account the presence of the coupled channels $j=$~\LiPiz, \SPi\, (\SipPim, \SimPip, \SizPiz), and \kzerobar n in the following way~\cite{Lednicky:1981su,Haidenbauer:2018jvl,Kamiya:2019uiw}:

\begin{align}\label{eq:corrfun_CC}
    C_{\km p}(\ks) & = \int d^3 r^* S_{\km p} (r^*) |\psi_{\km p} (\ks,r^*)|^2 + \sum_{j} \omega _j \int d^3 r^* S_j (r^*) |\psi_j (k^*,r^*)|^2\nonumber \\
    & = C_{\km p} ^{\rm el.} (\ks) + C_{\kzerobar n} ^{\rm inel.} (\ks) +C_{\SipPim} ^{\rm inel.} (\ks)+C_{\SimPip} ^{\rm inel.} (\ks)+C_{\SizPiz} ^{\rm inel.} (\ks)+C_{\LiPiz} ^{\rm inel.} (\ks).
\end{align}

\sloppy The first integral on the right-hand side of the equation describes the elastic processes \mbox{$\km p \xleftrightarrow{} \km p$} in which the initial state produced in the collision and the final measured pair are the same. The second integral in Eq.~(\ref{eq:corrfun_CC}) represents the explicit contributions stemming from inelastic processes \mbox{$(j=\LiPiz,~\SPi,~\kzerobar n) \rightarrow \km p$}, in which initial and final states are different but share the same quantum numbers. The functions $S_{\km p} (r^*)$ and $S_j (r^*)$ represent the emitting source profiles, as a function of the relative distance $r^*$ in the pair rest frame, for \km p pairs and for the $j$ inelastic channels, respectively. Details on the modeling of the source are available in Section~\ref{sec:source}.

One of the main ingredients needed to calculate $ C_{\km p}(\ks)$ are
the elastic $\psi_{\km p} (\ks,r^*)$ and the inelastic $\psi_j (k^{*},r^*)$ wave functions to be evaluated in a coupled channel approach by solving the multi-channel Schr\"odinger equation.

Since the coupled channel dynamics mostly acts at inter-particle distances of the order of 1~fm, the inelastic terms shown in Eq.~(\ref{eq:corrfun_CC}) should be relevant for femtoscopic measurements performed in small colliding systems like \pp, \pPb, peripheral and semi-peripheral Pb--Pb.
It has been shown that the probed source sizes in such small  systems are around 1~fm~\cite{ALICE:Source} and the explicit inclusion of the inelastic correlations is needed to reproduce the data~\cite{FemtoKp_pp, Kamiya:2019uiw,ALICE:pL,ALICE:BBarpp}. These inelastic terms can modify the shape and the strength of the total correlation function as shown in Refs.~\cite{Fabbietti:2020bfg,Kamiya:2019uiw,Haidenbauer:2018jvl}. Channels opening below the threshold, such as \LiPiz and \SPi, introduce an enhancement of the femtoscopic signal at low momenta and channels opening above threshold, such as \kzerobar n, lead to the appearance of a cusp-like structure at the corresponding \ks value~\cite{Kamiya:2019uiw,Haidenbauer:2018jvl}. 
These modifications and the corresponding coupled channel contributions become negligible when the correlation function is measured in large colliding systems, such as central and semi-central \PbPb collisions, for which the source can reach radii of about 5~fm and above. Under these circumstances, the correlation function is mostly driven by the elastic contribution, given by the first term in Eq.~(\ref{eq:corrfun_CC}).

To properly account for the coupled channel correlations, additional quantities are needed: the conversion weights $\omega_j$.
These conversion weights are related to the number of pairs in each inelastic channel $j$ originating from primary particle produced in the collision and can hence be defined as $\omega_j=\omega ^{\rm prod}_j$.
The yields of pairs in the $j$ channel used to compute $\omega ^{\rm prod}_j$ are obtained in this analysis by using the statistical thermal model implemented in the \textsc{Thermal-FIST} (TF) package~\cite{Vovchenko:2019pjl}. 
Since the sizes of the source in pp, \pPb, and semi-peripheral \PbPb are small, the canonical statistical model (CSM) with incomplete equilibration of strangeness as implemented in TF~\cite{Vovchenko:2019kes} is used in this study ($\gamma_\mathrm{s}$-CSM). The three parameters of the model (i.e. the chemical freeze-out temperature \temp, the strangeness saturation parameter $\gamma_\mathrm{s}$, and the volume at midrapidity d$V$/d$y$) are extracted from the parameterisation provided in Ref.~\cite{Vovchenko:2019kes} at the average multiplicity of charged particles ($\langle \dndeta \rangle$) measured by ALICE for the different collision systems, energies, and centrality intervals considered in this analysis~\cite{ALICE:2015qqj, ALICE:2012xs, ALICE:2015juo}. The values of the parameters and their estimated uncertainties are summarised in Table~\ref{tab:TFparameters}. The uncertainties assigned to the different parameters are evaluated starting from the systematic uncertainties assigned to the measured $\langle \dndeta \rangle$ and propagated through  the parameterisation. For fixed freeze-out parameters, the TF framework provides the yields of the primary particles composing the pairs in each channel $j$. 
The final amount of pairs in the $j$ channel, $N_j$, is given by the product between the primary yields of the particles in the considered pair. 

\begin{table}[!ht]
\caption{Parameters used in the $\gamma_\mathrm{s}$-CSM model of \textsc{Thermal-FIST}~\cite{Vovchenko:2019kes,Vovchenko:2019pjl} for different colliding systems and centrality intervals. The average \dndeta ($\EuScript{M}=\langle \dndeta \rangle$) and the associated systematic uncertainties corresponding to the various colliding systems are taken from Refs.~\cite{ALICE:2015qqj,ALICE:2012xs,ALICE:2015juo}. The total uncertainties assigned to \temp, $\gamma_\mathrm{s}$ and d$V$/d$y$ are obtained as described in the text. 
}

\begin{center}
\begin{tabular}{|l|l|c|c|l|}
\hline
System & $\EuScript{M}$ & \temp(MeV) & $\gamma_\mathrm{s}$ & d$V$/d$y$ (fm$^3$) \\
\hline

pp, \s=13~TeV & 6.94$^{+0.10}_{-0.08}$     & 171$\pm$1 & 0.78$\pm$0.06 & 16.66$\pm$1.39  \\ \hline\hline
\pPb, 0--20\% & 35.42$\pm$1.44	& 167$\pm$1 & 0.86$\pm$0.33 & 85.01$\pm$7.08  \\ \hline
\pPb, 20--40\% & 23.12$\pm$0.52	& 168$\pm$1 & 0.83$\pm$0.20 & 55.49$\pm$4.62  \\ \hline
\pPb, 40--100\% & 9.88$\pm$0.42 & 170$\pm$1 & 0.79$\pm$0.09 & 23.71$\pm$1.98  \\ \hline\hline
\PbPb, 60--70\% & 96.3$\pm$5.8 & 164$\pm$1 & 0.95$\pm$0.59 & 231.12$\pm$19.26\\ \hline
\PbPb, 70--80\% & 44.9$\pm$3.4 & 166$\pm$1 & 0.88$\pm$0.43 & 107.76$\pm$8.98 \\ \hline
\PbPb, 80--90\% & 17.52$\pm$1.89 & 169$\pm$1 & 0.81$\pm$0.15 & 42.05$\pm$3.50 \\ \hline
\end{tabular}
\end{center}
\label{tab:TFparameters}
\end{table}

The kinematic distributions of the produced pairs are obtained with toy Monte Carlo simulations using a Blast-Wave (BW) parameterisation~\cite{Schnedermann19932462} of the \pt of each particle composing the pair of interest. The parameters of the BW are taken from Ref.~\cite{pPbdata} for \pPb data and from Ref.~\cite{PbPbdata} for \PbPb data, and are assumed to describe also the spectra of particles not measured by ALICE (e.g. neutrons and $\Sigma$s). The parameters used for the most peripheral \pPb centrality intervals are also used to describe the spectra of particles produced in pp collisions at~\onethree. Once the momentum distributions of the two particles are generated, only pairs in the $j$ channel with relative momentum \ks below 200 \MeVc were selected and taken into consideration. The obtained distribution is then integrated over transverse-momentum space and the BW yields are extracted. These yields, for the \km p and different $j$-pairs, are finally rescaled by the primary TF $N_j$ results in order to account for the proper produced abundances in each colliding system.
The corresponding production weights $\omega ^{\rm prod}_j$ are obtained by dividing these final yields with the total production of \km p pairs, considered as the reference for  
this study. The relative production weights $\omega ^{\rm prod}_j$ are reported in Table~\ref{tab:weightsppb} for the different centralities and collision systems. The systematic uncertainties on the $\omega ^{\rm prod}_j$ reported in the table were evaluated by varying the parameters in the $\gamma_\mathrm{s}$-CSM model implemented in TF ($\sigma^{\rm TF}$) by $\pm1\sigma^{\rm TF}$ and by varying the upper limit of the relative momenta \kstar of the produced pairs by $\pm100$~\MeVc. The latter variation represents the dominant contribution to the uncertainty for pairs produced below the \km p threshold. The systematic uncertainty assigned to the $\omega ^{\rm prod}_j$ is around 20\% for \km p and \kzerobar n pairs and around 50\% for the other coupled channel pairs. Since the productions of the three different species in the isospin triplet of $\uppi$ and $\Sigma$ particles are similar in TF, in the following analysis a single $\omega ^{\rm prod}_j$ for \SPi, evaluated as the average of the three channels, will be used.

\begin{table}[!ht]
\caption{Production weights $\omega ^{\rm prod}_j$ for $j=\LiPiz,~\SPi,~\kzerobar n$ pairs for minimum bias pp collisions at \onethree and for three different centrality intervals in \pPb and \PbPb collisions at \fivenn. The average \dndeta ($\EuScript{M}=\langle \dndeta \rangle$) and the associated systematic uncertainties used to compute the production yield of each species, as described in Section~\ref{sec:ck_theo}, are also reported in Refs.~\cite{ALICE:2015qqj,ALICE:2012xs,ALICE:2015juo}. For the calculation, an upper limit of \ks of~200~\MeVc for the pairs was imposed. The systematic uncertainties associated to each $\omega ^{\rm prod}_j$ are evaluated as described in Section~\ref{sec:ck_theo}.
}
\begin{center}
\begin{adjustbox}{width=\textwidth}
\begin{tabular}{|l|c|c|c|c|c|c|c|}
\hline
\multirow{3}{*}{Pairs}  & \multicolumn{1}{c|}{pp}  & \multicolumn{3}{c|}{\pPb}        & \multicolumn{3}{c|}{\PbPb}          \\ \cline{2-8} 
                       & \onethree, MB      & 0--20\% & 20--40\% & 40--100\% & 60--70\% & 70--80\% & 80--90\% \\ \cline{2-8} 
   & $\EuScript{M}$ = $6.94^{+0.10}_{-0.08}$ & $\EuScript{M}$ = $35.42\pm 1.44$ & $\EuScript{M}$ = 23.12$\pm 0.52$ & $\EuScript{M}$ = 9.88$\pm 0.42$ &  $\EuScript{M}$ = 96.3$\pm 5.8$ &  $\EuScript{M}$ = 44.9$\pm 3.4$ &  $\EuScript{M}$ = 17.52$\pm 1.89$ \\ \hline

\km p       &  1.00  & 1.00  & 1.00   &  1.00   &  1.00  & 	1.00  & 1.00   \\ \hline
\kzerobar n &  0.97 $\pm$ 0.20 & 0.98 $\pm$  0.20 & 0.97  $\pm$  0.20 &  0.99  $\pm$  0.20 &  0.99  $\pm$  0.20 &  0.99 $\pm$  0.20 & 0.99 $\pm$ 0.20   \\ \hline
\SipPim     &  1.41 $\pm$ 0.70 & 1.41 $\pm$  0.70 & 1.35  $\pm$  0.67 &  1.27  $\pm$  0.63 &  1.46  $\pm$  0.73 &  1.38 $\pm$  0.69 &	1.30 $\pm$ 0.65	  \\ \hline
\SimPip     &  1.42 $\pm$ 0.71 & 1.42 $\pm$  0.71 & 1.35  $\pm$  0.67 &  1.29  $\pm$  0.64 &  1.47  $\pm$  0.73 &  1.39 $\pm$  0.69 &	1.31 $\pm$ 0.65	  \\ \hline
\SizPiz     &  1.37 $\pm$ 0.68 & 1.41 $\pm$  0.70 & 1.38  $\pm$  0.70 &  1.22  $\pm$  0.61 &  1.46  $\pm$  0.73 &  1.38 $\pm$  0.69 &	1.31 $\pm$ 0.65   \\ \hline
\LiPiz      &  1.96 $\pm$ 0.93 & 2.07 $\pm$  1.03 & 1.96  $\pm$  0.93 &  1.86  $\pm$  0.93 &  1.48  $\pm$  0.74 &  1.40 $\pm$  0.70 &	1.32 $\pm$ 0.66	  \\ \hline

\end{tabular}
\end{adjustbox}
\end{center}
\label{tab:weightsppb}
\end{table}

\subsection{Experimental characterization of the emitting source \Sr}\label{sec:source}

A fundamental ingredient entering in the evaluation of the theoretical correlation function in Eq.~(\ref{eq:corrfun_CC}) is the emitting source for the elastic part \Sr and for the inelastic terms $S_j(r^*)$.
Recently, a data-driven analysis based on p--p correlations measured in \pp collisions provided a model for the emitting source of baryon--baryon pairs in small colliding systems~\cite{ALICE:Source}.
The source is composed of two components: a Gaussian core with a common radius $r_{\rm core}$, which scales with the transverse mass \mt of the pair due to possible collective effects (e.g. radial flow) and an exponential tail stemming from short-lived resonances \mbox{($c\tau \approx $1--2\,fm)} strongly decaying into the particles composing the pair of interest.
In the baryon--baryon sector, the determination of $r_{\rm core}$ is anchored to the p--p correlation since its underlying strong interaction is the best known. %in the baryon--baryon sector
The emitting source obtained with this model was used in several baryon--baryon and baryon--antibaryon femtoscopic measurements~\cite{ALICE:pOmega,ALICE:pSig0,ALICE:BBarpp} to study the underlying strong interaction.   

In the meson--baryon sector the role played by the p--p interaction in constraining the source for \mbox{baryon--baryon} pairs is overtaken in this study by the \kp p system since the interaction is well known~\cite{AokiJidoKplusp} and coupled channels are not present~\cite{FemtoKp_pp}. 
The influence of short-lived resonances ($c\tau < 5$\,fm) on the source is quantified by evaluating the yields of each resonance with TF~\cite{Vovchenko:2019pjl} and by extracting the decay kinematics using transport model dynamics implemented in EPOS~\cite{EPOS}. 
According to these calculations, which entail the full decay chain from heavy to light particles, around 52\% (36\%) of the total \kp (p) yield is primordial. The relevant contributions of short-lived resonances decaying into \kp are summarised in Table~\ref{tab:ThermalFIST_resoYields_chargedkaons}. For these calculations, every hadron consisting of light and strange quarks was taken into account. As already introduced in Section~\ref{sec:CFanalysis}, due to its large lifetime of 46~fm, the $\phi$ (1020) is considered in this analysis as a primary particle, and its contribution is taken into account as done for secondary particles.  The resonance yields for the proton are the same as reported in Ref.~\cite{ALICE:Source}.

\begin{table}[!ht]
\begin{center}    
\caption{List of resonances contributing at least 1\% to the yield of $\rm{K^{+}}$. These fractions are computed with \textsc{Thermal-FIST} for \pp minimum bias collisions at \onethree, and are used also for the other collision systems.}

\begin{tabular}{|l |c|}
\hline
Resonances & Fraction ($\%$) \\
\hline 
$\rm{K^*(892)^0}$ & 21 \\
$\rm{K^*(892)^+}$ & 11 \\
$\rm{a_0(980)^+}$ & 1 \\
$\rm{K^{*}_{2}(1430)^0}$ & 1 \\
$\rm{K^{*}_{1}(1270)^0}$ & 1 \\
$\phi$ (1020)  & 6 \\
\hline
\end{tabular}
\label{tab:ThermalFIST_resoYields_chargedkaons}
\end{center}
\end{table}

The weighted average of the lifetimes of the resonances feeding into \kp (p) is 3.66~fm/$c$ (1.65~fm/$c$), while the weighted average of the masses is 1.05~\GeVmass (1.36~\GeVmass). The decay kinematics are extracted from the EPOS transport model by generating high-multiplicity pp events at $\s = 13$ TeV and selecting primordial \kp (p) and the resonances feeding into the particle of interest. The particle resonance cocktail is required to reproduce the weighted average of the masses (lifetimes) of the resonances feeding into \kp or p. Using as an input the fraction of primordial \kp and p, the weighted average of the lifetimes/masses and the corresponding decay kinematics, the source is modeled employing a dedicated Monte Carlo procedure, details of which are described in Ref.~\cite{ALICE:Source}.

In order to extract the core source size, the genuine \kp p correlation function $C(k^*)$ can be modeled  using the following equation

\begin{equation}
C(k^*) = \EuScript{N} \times (C_{\kp p}(k^*)), 
\label{eq:cf1}
\end{equation}

where the normalization constant, $\EuScript{N}$, is a free parameter of the fit and is introduced to take into account any remaining  contributions which survive the correction procedure described in Section~\ref{sec:ck_exp}. 

The correlation function $C(k^*)$ is fitted with Eq.~(\ref{eq:cf1}) in the range $0 < \ks < 250$~\MeVc. A variation of $\pm30$~\MeVc for the upper \kstar value is applied and accounted for in the systematic uncertainties. 
The theoretical correlation function, $C_{\kp p}(k^*)$, is evaluated using the CATS framework~\cite{Mihaylov:2018rva}. The strong potential for the \kp p is constructed based on the scattering amplitude in chiral SU(3) dynamics~\cite{AokiJidoKplusp}, using the procedure developed in Ref.~\cite{Miyahara:2018onh}. The Coulomb interaction in the same-charge pair is taken into account in the final potential directly using CATS. The source is modeled including the effects of short-lived resonances via a dedicated Monte Carlo procedure implemented in CATS, details are given in Ref.~\cite{ALICE:Source}. In order to account for the uncertainty related to the correction procedure, the $\lambda$ parameter estimation, and the baseline, the data were fitted several times. Each time either the $\lambda$ parameter or the parameters of the baseline defined in Eq.~(\ref{eq:unfoldinge}) were varied by $\pm1\sigma$. The radii obtained from the fits which minimise the $\chi^2$ per number of degrees of freedom ($\chi^2$/NDF)  represent the \rcore for \kp p pairs. 
In order to evaluate the systematic uncertainty on \rcore, half of the maximum variations among the different radii extracted from the multiple fits are considered.  
Possible biases in the input hadronic spectrum were cross checked by comparing the  resonance yields obtained from TF with model calculations based on Ref.~\cite{Becattini_2011}. The yields obtained with both methods agree within~5\%, for the short-lived as well as the long-lived resonances. Therefore, in order to account properly for discrepancies due to the model choice, the fraction of resonances that decay into K and p was varied by $\pm$10\%. Comparing the two different models, the extracted values for the weighted average of the lifetimes (masses) were found to be consistent within 10\%. 
Finally, the \ks cutoff in the EPOS transport code, which is used to extract the decay kinematics, is varied from 200~\MeVc to 150 and 250~\MeVc. The limit in \ks is used to accept pairs lying within the kinematic range for which femtoscopic effects are expected. All the variations  described  here are bootstrapped~\cite{Bootstrap} to obtain a systematic uncertainty related to resonance contributions. 

In Fig.~\ref{fig:KppCFfitpp}, the results obtained for the measured \kp p correlation in \pp collisions at \onethree are shown. The data are fitted with Eq.~(\ref{eq:cf1}) and a similar fitting procedure has been performed in \pPb collisions (see Fig.~\ref{fig:KppCFfitpPb}) and \PbPb collisions (see Fig.~\ref{fig:KppCFfitPbPb}). The data in all three colliding systems are well reproduced by the assumed \kp p interaction. 
The red band in the upper panels represents the model calculation and its systematic uncertainty. 
The lower panel of each figure shows the difference between data and model evaluated in the middle of each \kstar\ interval, and divided by the statistical uncertainty of data (n$_{\sigma_{\rm stat}}$). 
The width of the bands represents the n$_{\sigma_{\rm stat}}$ range associated to the model systematic variations. The reduced $\chi^2$ are also shown.
The values of $r_\mathrm{core}$ are reported on the figures along with the corresponding effective Gaussian source size $r_{\rm eff}$, which describes the resulting core source with the resonance contributions. The associated statistical and systematic uncertainties are also reported. 
This study shows an agreement between the results obtained from p--p correlations in Ref.~\cite{ALICE:Source} and the extracted \rcore for \kp p pairs and the \kp p system is used to fix the core radius for both the elastic channel $S_{\km p} (r^*)$ and the inelastic channels  $S_j (r^*)$ of the \km p interaction as described in Eq.~(\ref{eq:corrfun_CC}).

The sources of other pairs involving kaons and nucleons, such as $S_{\rm{K^- p}}(\rs)$ and $S_{\rm{\kzerobar n}}(\rs)$, are modeled assuming the same feed-down from resonances and decay kinematics as described for the \kp p pairs.
The emitting source for the inelastic correlations containing pions, $S_{\SipPim}(\rs),~S_{\SimPip}(\rs),~S_{\SizPiz}(\rs)$, and $S_{\LiPiz}(\rs)$, are treated separately from the source of the elastic \km p and charge-exchange \kzerobar n terms due to the presence of pions~\cite{FemtoKp_pp}.
The resonance contributions to the $\uppi$ part of $S_{\SipPim}(\rs),~S_{\SimPip}(\rs),~S_{\SizPiz}(\rs)$, and $S_{\LiPiz}(\rs)$ are assumed to be equivalent.
The feed-down from resonances to $\Sigma$ and $\Lambda$ baryons is similar, hence the sources for the \SPi and \LiPiz contributions are modeled using resonances and decay kinematics related to $\Lambda$ particles as in Ref.~\cite{ALICE:Source}, together with the ones necessary for the pions. Around 28\% (36\%) of the total $\uppi$ ($\Sigma$/$\Lambda$) yields are primordial. The contributions to the $\uppi$ yield stemming from the different resonances are summarised in Table~\ref{tab:ThermalFIST_resoYields_chargedpions}.
The remaining resonances, not explicitly quoted in Table~\ref{tab:ThermalFIST_resoYields_chargedpions}, contribute at the sub percent level and largely decay into a nucleon and a $\uppi$. Around 12\% (including the yield from $\omega$ (782)) of the total $\uppi$ yield stems from strongly decaying resonances with a lifetime larger than 5~fm/$c$. Due to the long lifetime, these contributions are included in the weak feed-down embedded in the $\lambda$ parameters, consistent with the handling of the \kp stemming from the $\phi$ decay.

\begin{table}[!ht]
\begin{center}    
\caption[Resonances contributions for $\uppi$]{List of resonances which contribute at least 1\% to the yield of $\uppi^+$. The fractions are computed with \textsc{Thermal-FIST} for \pp minimum bias collisions at \onethree, and used also for the other collision systems.}

\begin{tabular}{|l |c|}
\hline
Resonances & Fraction (\%) \\
\hline 
$\rm{\rho(770)^0}$ & {9.0} \\
$\rm{\rho(770)^+}$ & {8.7} \\
$\rm{\omega}$({782}) & {7.7} \\
$\rm{K^{*}({892})^+}$  & {2.3} \\
$\rm{\overline{K}^*({892})^0}$  & {2.6} \\
$\rm{b_1({1235})^0}$  & {1.9} \\
$\rm{a_2({1320})^+}$  & {1.5} \\
$\rm{\eta}$ & {1.5} \\
$\rm{a_1({1260})^+}$  & {1.4} \\
$\rm{f_2({1270})}$  & {1.4} \\
$\rm{a_0({980}) ^+}$ & {1.4} \\
$\rm{h_1({1170})}$  & {1.2} \\
\hline
\end{tabular}
\label{tab:ThermalFIST_resoYields_chargedpions}
\end{center}
\end{table}

The weighted average of the lifetimes of the resonances feeding into $\uppi$ ($\Sigma$/$\Lambda$) is 1.5~fm/$c$ (4.7~fm/$c$), while the weighted average of the masses is 1.13~\GeVmass (1.46~\GeVmass).
The decay kinematics are extracted by using the transport model dynamics implemented in EPOS, as done for \kp. 

Following this procedure, the source terms for the coupled channel contributions as well as for the genuine \km p correlation in Eq.~(\ref{eq:corrfun_CC}) are evaluated by using the size of the Gaussian core source from the \kp p results as a starting value in the fit procedure and by taking into account the specific feed-down from resonances. 
Similarly to the \kp p modeling, Eq.~(\ref{eq:cf1}) is used to model the measured \km p correlation in the three colliding systems.  

\sloppy The theoretical correlation function, including the elastic and inelastic contributions, is evaluated within the CATS framework using the wave functions obtained from state-of-the-art chiral coupled channel~\mbox{\kbarN--\SPi--\LiPiz} potentials~\cite{Miyahara:2018onh} based on the scattering amplitudes in Refs.~\cite{Ikeda:2012au,Ikeda:2011pi} (hereafter chiral model). 
The boundary conditions on the outgoing \km p wave function properly take into account the contributions from the inelastic channels \kzerobar n, \SPi and \LiPiz~\cite{Kamiya:2019uiw}. The parameters of the chiral model are tuned to reproduce the available scattering data and the kaonic hydrogen constraints from the SIDDHARTA experiment~\cite{Bazzi:2011zj}.

\section{Results and discussion}\label{sec:Results}

The results for \km p pairs are shown in the upper panel of Fig.~\ref{fig:KmpCFfitpp}, Fig.~\ref{fig:KmpCFfitpPb} and Fig.~\ref{fig:KmpCFfitPbPb} for \pp, \pPb and \PbPb collisions, respectively. 
The effect of the coupled channels on the measured correlation function can be seen explicitly by comparing the data in the three colliding systems.
For the centrality interval \mbox{60--70}\% in \PbPb collisions (left panel in Fig.~\ref{fig:KmpCFfitPbPb}), in which the largest source size, considered in this study, is achieved, no clear evidence of the expected \kzerobar n opening cusp is observed in the data. The disappearance of this cusp in large colliding systems is expected as described in~\cite{Kamiya:2019uiw,Fabbietti:2020bfg} based on the coupled channel formulation. The lack of a cusp in the correlation function is also observed in \km p femtoscopic measurements performed in central \PbPb collisions~\cite{FemtoKp_PbPb} in which the emitting source size can exceed radii of 5~fm. As the source size decreases, moving to more peripheral \PbPb collisions (middle and right panels in Fig.~\ref{fig:KmpCFfitPbPb}), the cusp structure at $\ks \approx 59$~\MeVc becomes more evident. 
The opening of the \kzerobar n channel is noticed as well in the measured correlation function in \pp (Fig.~\ref{fig:KmpCFfitpp}) and in the three centrality classes in \pPb collisions (Fig.~\ref{fig:KmpCFfitpPb}), in which the core source size is around 1~fm. 

The results are compared with the theoretical predictions from the chiral potentials described in Section~\ref{sec:CFanalysis}. 
The blue bands visible in Figs.~\ref{fig:KmpCFfitpp},~\ref{fig:KmpCFfitpPb}, and~\ref{fig:KmpCFfitPbPb} are obtained by fixing the production weights $\omega^{\rm prod}_j$ to the expected values in Table~\ref{tab:weightsppb}. 
An agreement with the data is achieved in particular for the colliding systems with the larger source sizes as in the first panel of Fig.~\ref{fig:KmpCFfitPbPb}. As the source size decreases, moving to more peripheral \PbPb collisions, to \pPb and \pp systems, the theoretical band clearly underestimates the data in the region of the \kzerobar n cusp.
To quantify this deviation, a scaling factor $\alpha_j$ is introduced in the definition of the conversion weights $\omega_j = \alpha_j \times \omega^{\rm prod}_j$. 
The red bands in Figs.~\ref{fig:KmpCFfitpp},~\ref{fig:KmpCFfitpPb}, and~\ref{fig:KmpCFfitPbPb} are obtained by letting $\alpha_j$ free for the \kzerobar n and \SPi contributions,  while keeping the production weights fixed. The scaling factor corresponding to the \LiPiz channel is kept at unity due to the negligible coupling of this channel to the \km p system~\cite{LPiCoupling1,LPiCoupling2,Haidenbauer:2018jvl,Kamiya:2019uiw}. The bottom panels of Fig.~\ref{fig:KmpCFfitpp}, Fig.~\ref{fig:KmpCFfitpPb}, and Fig.~\ref{fig:KmpCFfitPbPb} show the \mbox{data-to-model} difference expressed in terms of number of statistical standard deviations ($\sigma_{\mathrm{stat}}$) of the data. The width of the bands represents the n$_{\sigma_{\rm stat}}$ range associated to the model systematic variations. Blue squares show the n$_{\sigma_{\mathrm{stat}}}$ distribution when the  $\alpha_j$ are fixed to unity, while red squares show the results when $\alpha_j$ for  $j=\kzerobar \rm n$,~\SPi are free parameters of the fit. The $\chi^2$ over the number of degree of freedom ($\chi^2$/NDF) of the fit  obtained by using the two approaches are also shown. Clearly, the current tune of the chiral model is unable to properly describe the opening of the inelastic \kzerobar n channel.

\begin{figure}[!htb]  
\begin{center}
\includegraphics[width=0.6\textwidth]{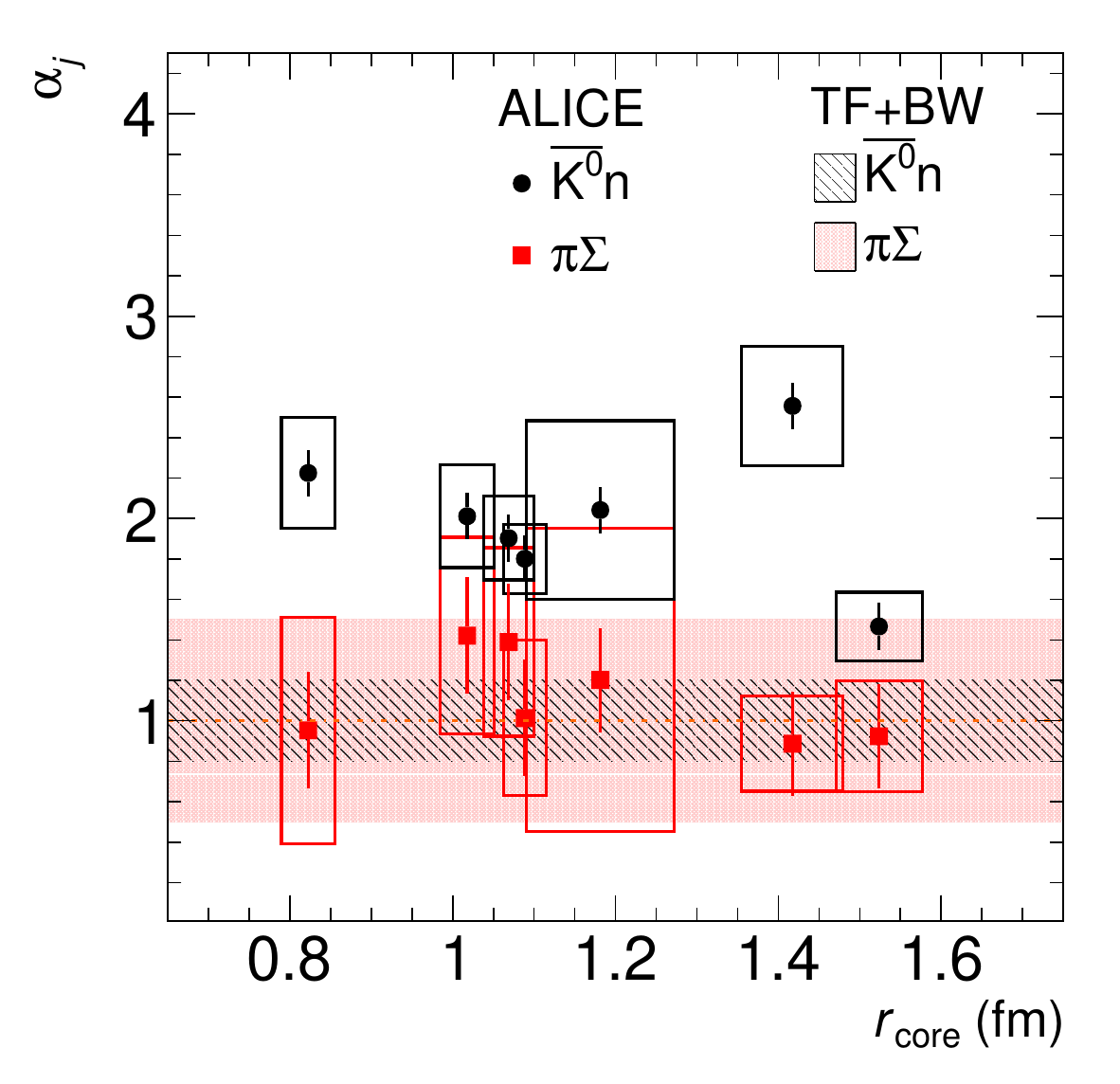}
\end{center}
\caption{Scaling factor ($\alpha_j $) for \kzerobar n (black circles) and \SPi (red squares) extracted from the different fits of the \km p 
correlation function as a function of the core radius $r_{\mathrm{core}}$ extracted for  pp, \pPb and \PbPb collisions. 
The vertical error bars and boxes represent the statistical and systematic uncertainties on the extracted parameters, respectively. The widths of the boxes represent the systematic uncertainty associated to each extracted $r_{\mathrm{core}}$. The black and red bands represent the uncertainty coming from the yield estimates in TF and the variations applied in the BW kinematics summed in quadrature as described in the text for \kzerobar n and \SPi, respectively.}
\label{fig:Weightskmp}
\end{figure}

The results of the extracted $\alpha_j$ for $j=\kzerobar \rm n$,~\SPi are shown in Fig.~\ref{fig:Weightskmp} as a function of the \rcore radius, for all colliding systems. The black and red bands represent the uncertainties coming from the yield estimates in TF and the variations applied in the BW kinematics for \kzerobar n and \SPi production weights, respectively. 
The vertical error bars in the figure represent the statistical uncertainty of the extracted parameter, while the boxes represent the systematic uncertainty on the conversion weights obtained by repeating the fit several times and by varying within statistical and systematic uncertainties \rcore  
and the amount of resonances that contribute to the source distribution. 
The \SPi scaling factors (red squares) obtained at different core radii are compatible with unity, indicating that this coupling to \km p is well modeled by the underlying chiral interaction.
The extracted scaling factor for the \kzerobar n coupling (black circles)  significantly deviates from unity for values of \rcore below 1.5~fm. 

As can be seen in Fig.~\ref{fig:Weightskmp}, for \PbPb collisions at 60--70\% centrality ($r_\mathrm{{core}}$ = $1.53\pm0.05\pm0.24$~fm), the scaling factor $\alpha _{\kzerobar \rm n}$ is consistent with unity. Both fit approaches shown in the left panel of Fig.~\ref{fig:KmpCFfitPbPb} can be used to describe the data, as can be seen in the $\rm n_{\sigma_{\rm stat}}$ distribution. These results are expected since the \mbox{60--70\%} centrality corresponds to the largest source size ($\rcore = 1.53$~fm) considered in this work, in which the coupled channel effects on the correlation are negligible.
As the source decreases and as the \kzerobar n cusp appears, the fit assuming free scaling factors provides a better description of the data and the extracted $\alpha_{\kzerobar n}$ are larger than unity. This deviation is likely due to the fact that a direct experimental constraint of the \km p to \kzerobar n coupling was not available before the measurements presented here and for the similar analysis performed in minimum bias events in \pp collisions at $\sqrt{s}=5.02$,~7, and 13~TeV~\cite{FemtoKp_pp}. 
The deviation from unity of $\alpha _{\kzerobar \rm n}$ indicates that the transition between the \km p and the \kzerobar n channel currently implemented in the chiral model model is too weak. Since other observables are also affected by the coupling of \km p and \kzerobar n, it is necessary in future studies to update the $S=-1$ meson-baryon scattering amplitude of  \kbarN--\SPi--\LPi system by including the present correlation function measurements in addition to the available kaonic hydrogen and scattering data. Moreover, the observed strong negative correlation between the coupling weights $\alpha_{\kzerobar n}$ and  $\alpha_{\SPi}$ indicates that a revision of the full coupled-channel \km p potential is required to describe the measured correlation function.

The data presented in this work provide unique constraints to pin down the coupling strength to the \kzerobar n channel and indicate the first difference between the chiral model and experimental data on the \km p interaction. A fine tuning of the chiral model is beyond the scope of this article since it requires a more detailed investigation on how much variation of the two isospin components  of the \kbarN interaction is allowed in order to still fulfil the scattering data and kaonic hydrogen constraints,  above and below threshold, respectively.

\section{Summary}\label{sec:Summary}

This study contains a detailed and comprehensive analysis of the \kp p 
and \km p 
correlation function in different colliding systems.  In the \km p correlation function, it is possible to observe the opening of the \kzerobar n channel in \pp, \pPb, and peripheral \PbPb collisions ($r_\mathrm{core}\lesssim$~1.5~fm), while it is less pronounced in \mbox{semi-central} \PbPb collisions, hence for larger distances.
By using the strong potential for \kp p pairs based on scattering amplitudes in chiral SU(3) dynamics and the Coulomb potential, it was possible to extract the common core radius ($r_{\rm core}$) of the Gaussian source, which, together with an exponential tail, is used to model the particle emitting source of \kp p. 
The same $r_{\rm core}$ is used to model the source for \km p. 
In this case, the source is composed of the elastic \km p  and the charge-exchange \kzerobar n terms ($S_{\rm KN}$), and the inelastic correlations containing a pion term ($S_{\SPi/\LiPiz}$). The accurate source description allows us to study the coupling strength associated with each of the inelastic channels as a function of the source size. In order to assess the $\omega_j$ conversion parameters of the different coupled channels present in the \km p system, a novel approach was used to estimate the contributions of the inelastic channels to the measured correlation function. The $\omega_j$ parameters can be expressed as the product of two terms: $\omega_j ^{\rm prod}$, which takes into consideration the production yield of each particle pairs, and the scaling factor $\alpha_j$. The latter should be equal to unity if the coupling strength is correctly estimated within the Kyoto model. 
From the fits to the measured correlation functions with the state-of-the-art Kyoto model, calculated within the coupled channel approach, it is possible to observe that the dynamics of the coupled channels is under control in the case of \SPi, while the deviation from unity of $\alpha_{\kzerobar \rm n}$ indicates that the transition between the \km p and the \kzerobar n channel, as currently implemented in the Kyoto model, is too weak. 
Hence, the data presented in this work provide a unique constraint to pin down the coupling strength to the \kzerobar n channel.

%%%%%%%%%%%%%%%%%%%%%%%%%%%%%%%%
% end main text 
%%%%%%%%%%%%%%%%%%%%%%%%%%%%%%%%
\newpage
%%%%% acknowledgements - handled by EB chairs 
\newenvironment{acknowledgement}{\relax}{\relax}
\begin{acknowledgement}
\section*{Acknowledgements}
% add specific acknowledgements here 
% ...but please don't remove the line below: funding agencies
% will be acknowledged with a custom tex file handled by EB chairs after Collab Round 2
% Version: 2022-05-07

The ALICE Collaboration would like to thank all its engineers and technicians for their invaluable contributions to the construction of the experiment and the CERN accelerator teams for the outstanding performance of the LHC complex.
The ALICE Collaboration gratefully acknowledges the resources and support provided by all Grid centres and the Worldwide LHC Computing Grid (WLCG) collaboration.
The ALICE Collaboration acknowledges the following funding agencies for their support in building and running the ALICE detector:
A. I. Alikhanyan National Science Laboratory (Yerevan Physics Institute) Foundation (ANSL), State Committee of Science and World Federation of Scientists (WFS), Armenia;
Austrian Academy of Sciences, Austrian Science Fund (FWF): [M 2467-N36] and Nationalstiftung f\"{u}r Forschung, Technologie und Entwicklung, Austria;
Ministry of Communications and High Technologies, National Nuclear Research Center, Azerbaijan;
Conselho Nacional de Desenvolvimento Cient\'{\i}fico e Tecnol\'{o}gico (CNPq), Financiadora de Estudos e Projetos (Finep), Funda\c{c}\~{a}o de Amparo \`{a} Pesquisa do Estado de S\~{a}o Paulo (FAPESP) and Universidade Federal do Rio Grande do Sul (UFRGS), Brazil;
Bulgarian Ministry of Education and Science, within the National Roadmap for Research Infrastructures 2020¿2027 (object CERN), Bulgaria;
Ministry of Education of China (MOEC) , Ministry of Science \& Technology of China (MSTC) and National Natural Science Foundation of China (NSFC), China;
Ministry of Science and Education and Croatian Science Foundation, Croatia;
Centro de Aplicaciones Tecnol\'{o}gicas y Desarrollo Nuclear (CEADEN), Cubaenerg\'{\i}a, Cuba;
Ministry of Education, Youth and Sports of the Czech Republic, Czech Republic;
The Danish Council for Independent Research | Natural Sciences, the VILLUM FONDEN and Danish National Research Foundation (DNRF), Denmark;
Helsinki Institute of Physics (HIP), Finland;
Commissariat \`{a} l'Energie Atomique (CEA) and Institut National de Physique Nucl\'{e}aire et de Physique des Particules (IN2P3) and Centre National de la Recherche Scientifique (CNRS), France;
Bundesministerium f\"{u}r Bildung und Forschung (BMBF) and GSI Helmholtzzentrum f\"{u}r Schwerionenforschung GmbH, Germany;
General Secretariat for Research and Technology, Ministry of Education, Research and Religions, Greece;
National Research, Development and Innovation Office, Hungary;
Department of Atomic Energy Government of India (DAE), Department of Science and Technology, Government of India (DST), University Grants Commission, Government of India (UGC) and Council of Scientific and Industrial Research (CSIR), India;
National Research and Innovation Agency - BRIN, Indonesia;
Istituto Nazionale di Fisica Nucleare (INFN), Italy;
Japanese Ministry of Education, Culture, Sports, Science and Technology (MEXT) and Japan Society for the Promotion of Science (JSPS) KAKENHI, Japan;
Consejo Nacional de Ciencia (CONACYT) y Tecnolog\'{i}a, through Fondo de Cooperaci\'{o}n Internacional en Ciencia y Tecnolog\'{i}a (FONCICYT) and Direcci\'{o}n General de Asuntos del Personal Academico (DGAPA), Mexico;
Nederlandse Organisatie voor Wetenschappelijk Onderzoek (NWO), Netherlands;
The Research Council of Norway, Norway;
Commission on Science and Technology for Sustainable Development in the South (COMSATS), Pakistan;
Pontificia Universidad Cat\'{o}lica del Per\'{u}, Peru;
Ministry of Education and Science, National Science Centre and WUT ID-UB, Poland;
Korea Institute of Science and Technology Information and National Research Foundation of Korea (NRF), Republic of Korea;
Ministry of Education and Scientific Research, Institute of Atomic Physics, Ministry of Research and Innovation and Institute of Atomic Physics and University Politehnica of Bucharest, Romania;
Ministry of Education, Science, Research and Sport of the Slovak Republic, Slovakia;
National Research Foundation of South Africa, South Africa;
Swedish Research Council (VR) and Knut \& Alice Wallenberg Foundation (KAW), Sweden;
European Organization for Nuclear Research, Switzerland;
Suranaree University of Technology (SUT), National Science and Technology Development Agency (NSTDA), Thailand Science Research and Innovation (TSRI) and National Science, Research and Innovation Fund (NSRF), Thailand;
Turkish Energy, Nuclear and Mineral Research Agency (TENMAK), Turkey;
National Academy of  Sciences of Ukraine, Ukraine;
Science and Technology Facilities Council (STFC), United Kingdom;
National Science Foundation of the United States of America (NSF) and United States Department of Energy, Office of Nuclear Physics (DOE NP), United States of America.
In addition, individual groups or members have received support from:
Marie Sk\l{}odowska Curie, European Research Council, Strong 2020 - Horizon 2020 (grant nos. 950692, 824093, 896850), European Union;
Academy of Finland (Center of Excellence in Quark Matter) (grant nos. 346327, 346328), Finland;
Programa de Apoyos para la Superaci\'{o}n del Personal Acad\'{e}mico, UNAM, Mexico.
\end{acknowledgement}

%%%%%%%% Bibliography 
\bibliographystyle{utphys}   % Remember we use title in the biblio
\bibliography{bibliography}
%\input {bibliography.tex}  

%%%%%%%%%%%%%%%%%%%%%%%%%%%%%%%%
% Appendices: yours (if any) + authorlist
%%%%%%%%%%%%%%%%%%%%%%%%%%%%%%%%
\newpage
\appendix

%
%\input{} % put your appendices here (if any)
%

%%%%% Authorlist - please do not touch: handled by EB chairs 
\section{The ALICE Collaboration}
\label{app:collab}
%ALICE Collaboration author list for 2022-05-07
\begingroup
\small
\begin{flushleft}

S.~Acharya\,{\orcidlink{0000-0002-9213-5329}}\,\Irefn{org125}\textsuperscript{,}\Irefn{org132}\And
D.~Adamov\'{a}\,{\orcidlink{0000-0002-0504-7428}}\,\Irefn{org86}\And
A.~Adler\Irefn{org69}\And
G.~Aglieri~Rinella\,{\orcidlink{0000-0002-9611-3696}}\,\Irefn{org32}\And
M.~Agnello\,{\orcidlink{0000-0002-0760-5075}}\,\Irefn{org29}\And
N.~Agrawal\,{\orcidlink{0000-0003-0348-9836}}\,\Irefn{org50}\And
Z.~Ahammed\,{\orcidlink{0000-0001-5241-7412}}\,\Irefn{org132}\And
S.~Ahmad\,{\orcidlink{0000-0003-0497-5705}}\,\Irefn{org15}\And
S.U.~Ahn\,{\orcidlink{0000-0001-8847-489X}}\,\Irefn{org70}\And
I.~Ahuja\,{\orcidlink{0000-0002-4417-1392}}\,\Irefn{org37}
% ...
\And
\renewcommand\labelenumi{\textsuperscript{\theenumi}~}

\section*{Affiliation notes}
\renewcommand\theenumi{\roman{enumi}}
\begin{Authlist}
\item \Adef{orgI} {Deceased}
\item \Adef{orgII} {Also at: Italian~National~Agency~for~New~Technologies,~Energy~and~Sustainable~Economic~Development~(ENEA),~Bologna,~Italy}
\item \Adef{orgIII} {Also at: Dipartimento~DET~del~Politecnico~di~Torino,~Turin,~Italy}
\item \Adef{orgVII} {Also at: Department of Physics, Tokyo Metropolitan University, Hachioji, Japan}
\item \Adef{orgVIII} {Also at: Helmholtz Institut f\"ur Strahlen- und Kernphysik and Bethe Center for Theoretical Physics, Universit\"at Bonn, Bonn, Germany}
\item \Adef{orgIV} {Also at: Department~of~Applied~Physics,~Aligarh~Muslim~University,~Aligarh,~India}
\item \Adef{orgV} {Also at: Institute~of~Theoretical~Physics,~University~of~Wroclaw,~Poland}
\item \Adef{orgVI} {Also at: University~of~Kansas,~Lawrence,~Kansas,~United~States}
\item \Adef{orgX} {Also at: An~institution~covered~by~a~cooperation~agreement~with~CERN}
\end{Authlist}

\section*{Collaboration Institutes}
\renewcommand\theenumi{\arabic{enumi}~}
\begin{Authlist}

\item \Idef{org1} A.I. Alikhanyan National Science Laboratory (Yerevan Physics Institute) Foundation, Yerevan, Armenia
\item \Idef{org2} AGH University of Science and Technology, Cracow, Poland
\item \Idef{org3} Bogolyubov Institute for Theoretical Physics, National Academy of Sciences of Ukraine, Kiev, Ukraine
\item \Idef{org4} Bose Institute, Department of Physics and Centre for Astroparticle Physics and Space Science (CAPSS), Kolkata, India
\item \Idef{org5} California Polytechnic State University, San Luis Obispo, California, United States
\item \Idef{org6} Central China Normal University, Wuhan, China
\item \Idef{org7} Centro de Aplicaciones Tecnol\'{o}gicas y Desarrollo Nuclear (CEADEN), Havana, Cuba
\item \Idef{org8} Centro de Investigaci\'{o}n y de Estudios Avanzados (CINVESTAV), Mexico City and M\'{e}rida, Mexico
\item \Idef{org9} Chicago State University, Chicago, Illinois, United States
\item \Idef{org10} China Institute of Atomic Energy, Beijing, China
\item \Idef{org11} Chungbuk National University, Cheongju, Republic of Korea
\item \Idef{org12} Comenius University Bratislava, Faculty of Mathematics, Physics and Informatics, Bratislava, Slovak Republic
\item \Idef{org13} COMSATS University Islamabad, Islamabad, Pakistan
\item \Idef{org14} Creighton University, Omaha, Nebraska, United States
\item \Idef{org15} Department of Physics, Aligarh Muslim University, Aligarh, India
\item \Idef{org16} Department of Physics, Pusan National University, Pusan, Republic of Korea
\item \Idef{org17} Department of Physics, Sejong University, Seoul, Republic of Korea
\item \Idef{org18} Department of Physics, University of California, Berkeley, California, United States
\item \Idef{org19} Department of Physics, University of Oslo, Oslo, Norway
\item \Idef{org20} Department of Physics and Technology, University of Bergen, Bergen, Norway
\item \Idef{org21} Dipartimento di Fisica, Universit\`{a} di Pavia, Pavia, Italy
\item \Idef{org22} Dipartimento di Fisica dell'Universit\`{a} and Sezione INFN, Cagliari, Italy
\item \Idef{org23} Dipartimento di Fisica dell'Universit\`{a} and Sezione INFN, Trieste, Italy
\item \Idef{org24} Dipartimento di Fisica dell'Universit\`{a} and Sezione INFN, Turin, Italy
\item \Idef{org25} Dipartimento di Fisica e Astronomia dell'Universit\`{a} and Sezione INFN, Bologna, Italy
\item \Idef{org26} Dipartimento di Fisica e Astronomia dell'Universit\`{a} and Sezione INFN, Catania, Italy
\item \Idef{org27} Dipartimento di Fisica e Astronomia dell'Universit\`{a} and Sezione INFN, Padova, Italy
\item \Idef{org28} Dipartimento di Fisica `E.R.~Caianiello' dell'Universit\`{a} and Gruppo Collegato INFN, Salerno, Italy
\item \Idef{org29} Dipartimento DISAT del Politecnico and Sezione INFN, Turin, Italy
\item \Idef{org30} Dipartimento di Scienze MIFT, Universit\`{a} di Messina, Messina, Italy
\item \Idef{org31} Dipartimento Interateneo di Fisica `M.~Merlin' and Sezione INFN, Bari, Italy
\item \Idef{org32} European Organization for Nuclear Research (CERN), Geneva, Switzerland
\item \Idef{org33} Faculty of Electrical Engineering, Mechanical Engineering and Naval Architecture, University of Split, Split, Croatia
\item \Idef{org34} Faculty of Engineering and Science, Western Norway University of Applied Sciences, Bergen, Norway
\item \Idef{org35} Faculty of Nuclear Sciences and Physical Engineering, Czech Technical University in Prague, Prague, Czech Republic
\item \Idef{org36} Faculty of Physics, Sofia University, Sofia, Bulgaria
\item \Idef{org37} Faculty of Science, P.J.~\v{S}af\'{a}rik University, Ko\v{s}ice, Slovak Republic
\item \Idef{org38} Frankfurt Institute for Advanced Studies, Johann Wolfgang Goethe-Universit\"{a}t Frankfurt, Frankfurt, Germany
\item \Idef{org39} Fudan University, Shanghai, China
\item \Idef{org40} Gangneung-Wonju National University, Gangneung, Republic of Korea
\item \Idef{org41} Gauhati University, Department of Physics, Guwahati, India
\item \Idef{org42} Helmholtz-Institut f\"{u}r Strahlen- und Kernphysik, Rheinische Friedrich-Wilhelms-Universit\"{a}t Bonn, Bonn, Germany
\item \Idef{org43} Helsinki Institute of Physics (HIP), Helsinki, Finland
\item \Idef{org44} High Energy Physics Group,  Universidad Aut\'{o}noma de Puebla, Puebla, Mexico
\item \Idef{org45} Horia Hulubei National Institute of Physics and Nuclear Engineering, Bucharest, Romania
\item \Idef{org46} Indian Institute of Technology Bombay (IIT), Mumbai, India
\item \Idef{org47} Indian Institute of Technology Indore, Indore, India
\item \Idef{org48} INFN, Laboratori Nazionali di Frascati, Frascati, Italy
\item \Idef{org49} INFN, Sezione di Bari, Bari, Italy
\item \Idef{org50} INFN, Sezione di Bologna, Bologna, Italy
\item \Idef{org51} INFN, Sezione di Cagliari, Cagliari, Italy
\item \Idef{org52} INFN, Sezione di Catania, Catania, Italy
\item \Idef{org53} INFN, Sezione di Padova, Padova, Italy
\item \Idef{org54} INFN, Sezione di Pavia, Pavia, Italy
\item \Idef{org55} INFN, Sezione di Torino, Turin, Italy
\item \Idef{org56} INFN, Sezione di Trieste, Trieste, Italy
\item \Idef{org57} Inha University, Incheon, Republic of Korea
\item \Idef{org58} Institute for Gravitational and Subatomic Physics (GRASP), Utrecht University/Nikhef, Utrecht, Netherlands
\item \Idef{org59} Institute of Experimental Physics, Slovak Academy of Sciences, Ko\v{s}ice, Slovak Republic
\item \Idef{org60} Institute of Physics, Homi Bhabha National Institute, Bhubaneswar, India
\item \Idef{org61} Institute of Physics of the Czech Academy of Sciences, Prague, Czech Republic
\item \Idef{org62} Institute of Space Science (ISS), Bucharest, Romania
\item \Idef{org63} Institut f\"{u}r Kernphysik, Johann Wolfgang Goethe-Universit\"{a}t Frankfurt, Frankfurt, Germany
\item \Idef{org64} Instituto de Ciencias Nucleares, Universidad Nacional Aut\'{o}noma de M\'{e}xico, Mexico City, Mexico
\item \Idef{org65} Instituto de F\'{i}sica, Universidade Federal do Rio Grande do Sul (UFRGS), Porto Alegre, Brazil
\item \Idef{org66} Instituto de F\'{\i}sica, Universidad Nacional Aut\'{o}noma de M\'{e}xico, Mexico City, Mexico
\item \Idef{org67} iThemba LABS, National Research Foundation, Somerset West, South Africa
\item \Idef{org68} Jeonbuk National University, Jeonju, Republic of Korea
\item \Idef{org69} Johann-Wolfgang-Goethe Universit\"{a}t Frankfurt Institut f\"{u}r Informatik, Fachbereich Informatik und Mathematik, Frankfurt, Germany
\item \Idef{org70} Korea Institute of Science and Technology Information, Daejeon, Republic of Korea
\item \Idef{org71} KTO Karatay University, Konya, Turkey
\item \Idef{org72} Laboratoire de Physique des 2 Infinis, Ir\`{e}ne Joliot-Curie, Orsay, France
\item \Idef{org73} Laboratoire de Physique Subatomique et de Cosmologie, Universit\'{e} Grenoble-Alpes, CNRS-IN2P3, Grenoble, France
\item \Idef{org74} Lawrence Berkeley National Laboratory, Berkeley, California, United States
\item \Idef{org75} Lund University Department of Physics, Division of Particle Physics, Lund, Sweden
\item \Idef{org76} Nagasaki Institute of Applied Science, Nagasaki, Japan
\item \Idef{org77} Nara Women{'}s University (NWU), Nara, Japan
\item \Idef{org78} National and Kapodistrian University of Athens, School of Science, Department of Physics , Athens, Greece
\item \Idef{org79} National Centre for Nuclear Research, Warsaw, Poland
\item \Idef{org80} National Institute of Science Education and Research, Homi Bhabha National Institute, Jatni, India
\item \Idef{org81} National Nuclear Research Center, Baku, Azerbaijan
\item \Idef{org82} National Research and Innovation Agency - BRIN, Jakarta, Indonesia
\item \Idef{org83} Niels Bohr Institute, University of Copenhagen, Copenhagen, Denmark
\item \Idef{org84} Nikhef, National institute for subatomic physics, Amsterdam, Netherlands
\item \Idef{org85} Nuclear Physics Group, STFC Daresbury Laboratory, Daresbury, United Kingdom
\item \Idef{org86} Nuclear Physics Institute of the Czech Academy of Sciences, Husinec-\v{R}e\v{z}, Czech Republic
\item \Idef{org87} Oak Ridge National Laboratory, Oak Ridge, Tennessee, United States
\item \Idef{org88} Ohio State University, Columbus, Ohio, United States
\item \Idef{org89} Physics department, Faculty of science, University of Zagreb, Zagreb, Croatia
\item \Idef{org90} Physics Department, Panjab University, Chandigarh, India
\item \Idef{org91} Physics Department, University of Jammu, Jammu, India
\item \Idef{org92} Physics Department, University of Rajasthan, Jaipur, India
\item \Idef{org93} Physics Program and International Institute for Sustainability with Knotted Chiral Meta Matter (SKCM2), Hiroshima University, Hiroshima, Japan
\item \Idef{org94} Physikalisches Institut, Eberhard-Karls-Universit\"{a}t T\"{u}bingen, T\"{u}bingen, Germany
\item \Idef{org95} Physikalisches Institut, Ruprecht-Karls-Universit\"{a}t Heidelberg, Heidelberg, Germany
\item \Idef{org96} Physik Department, Technische Universit\"{a}t M\"{u}nchen, Munich, Germany
\item \Idef{org97} Politecnico di Bari, Bari, Italy
\item \Idef{org98} Research Division and ExtreMe Matter Institute EMMI, GSI Helmholtzzentrum f\"ur Schwerionenforschung GmbH, Darmstadt, Germany
\item \Idef{org151} RIKEN iTHEMS, Wako, Japan
\item \Idef{org99} Saga University, Saga, Japan
\item \Idef{org100} Saha Institute of Nuclear Physics, Homi Bhabha National Institute, Kolkata, India
\item \Idef{org101} School of Physics and Astronomy, University of Birmingham, Birmingham, United Kingdom
\item \Idef{org102} Secci\'{o}n F\'{\i}sica, Departamento de Ciencias, Pontificia Universidad Cat\'{o}lica del Per\'{u}, Lima, Peru
\item \Idef{org103} Stefan Meyer Institut f\"{u}r Subatomare Physik (SMI), Vienna, Austria
\item \Idef{org104} SUBATECH, IMT Atlantique, Nantes Universit\'{e}, CNRS-IN2P3, Nantes, France
\item \Idef{org105} Suranaree University of Technology, Nakhon Ratchasima, Thailand
\item \Idef{org106} Technical University of Ko\v{s}ice, Ko\v{s}ice, Slovak Republic
\item \Idef{org107} The Henryk Niewodniczanski Institute of Nuclear Physics, Polish Academy of Sciences, Cracow, Poland
\item \Idef{org108} The University of Texas at Austin, Austin, Texas, United States
\item \Idef{org109} Universidad Aut\'{o}noma de Sinaloa, Culiac\'{a}n, Mexico
\item \Idef{org110} Universidade de S\~{a}o Paulo (USP), S\~{a}o Paulo, Brazil
\item \Idef{org111} Universidade Estadual de Campinas (UNICAMP), Campinas, Brazil
\item \Idef{org112} Universidade Federal do ABC, Santo Andre, Brazil
\item \Idef{org113} University of Cape Town, Cape Town, South Africa
\item \Idef{org114} University of Houston, Houston, Texas, United States
\item \Idef{org115} University of Jyv\"{a}skyl\"{a}, Jyv\"{a}skyl\"{a}, Finland
\item \Idef{org116} University of Kansas, Lawrence, Kansas, United States
\item \Idef{org117} University of Liverpool, Liverpool, United Kingdom
\item \Idef{org118} University of Science and Technology of China, Hefei, China
\item \Idef{org119} University of South-Eastern Norway, Kongsberg, Norway
\item \Idef{org120} University of Tennessee, Knoxville, Tennessee, United States
\item \Idef{org121} University of the Witwatersrand, Johannesburg, South Africa
\item \Idef{org122} University of Tokyo, Tokyo, Japan
\item \Idef{org123} University of Tsukuba, Tsukuba, Japan
\item \Idef{org124} University Politehnica of Bucharest, Bucharest, Romania
\item \Idef{org125} Universit\'{e} Clermont Auvergne, CNRS/IN2P3, LPC, Clermont-Ferrand, France
\item \Idef{org126} Universit\'{e} de Lyon, CNRS/IN2P3, Institut de Physique des 2 Infinis de Lyon, Lyon, France
\item \Idef{org127} Universit\'{e} de Strasbourg, CNRS, IPHC UMR 7178, F-67000 Strasbourg, France, Strasbourg, France
\item \Idef{org128} Universit\'{e} Paris-Saclay Centre d'Etudes de Saclay (CEA), IRFU, D\'{e}partment de Physique Nucl\'{e}aire (DPhN), Saclay, France
\item \Idef{org129} Universit\`{a} degli Studi di Foggia, Foggia, Italy
\item \Idef{org130} Universit\`{a} del Piemonte Orientale, Vercelli, Italy
\item \Idef{org131} Universit\`{a} di Brescia, Brescia, Italy
\item \Idef{org132} Variable Energy Cyclotron Centre, Homi Bhabha National Institute, Kolkata, India
\item \Idef{org133} Warsaw University of Technology, Warsaw, Poland
\item \Idef{org134} Wayne State University, Detroit, Michigan, United States
\item \Idef{org135} Westf\"{a}lische Wilhelms-Universit\"{a}t M\"{u}nster, Institut f\"{u}r Kernphysik, M\"{u}nster, Germany
\item \Idef{org136} Wigner Research Centre for Physics, Budapest, Hungary
\item \Idef{org137} Yale University, New Haven, Connecticut, United States
\item \Idef{org138} Yonsei University, Seoul, Republic of Korea
\item \Idef{org150} Yukawa Institute for Theoretical Physics, Kyoto University, Kyoto, Japan
\item \Idef{org139}  Zentrum  f\"{u}r Technologie und Transfer (ZTT), Worms, Germany
\item \Idef{org140} Affiliated with an institute covered by a cooperation agreement with CERN, 
\item \Idef{org141} Affiliated with an international laboratory covered by a cooperation agreement with CERN., 

\end{Authlist}
\endgroup

% Last update: 2022-05-07
  
\end{document}